\documentclass[
reprint,
superscriptaddress,
amsmath,
amssymb,
]{revtex4-2}
\usepackage[T1]{fontenc}

\usepackage{graphicx}
\usepackage{dcolumn}
\usepackage{bm}
\usepackage{xcolor}
\usepackage{afterpage}
\usepackage{multirow}
\usepackage{array}

\begin{document}

\preprint{APS/123-QED}

\title{Identifying Universal Spin Excitations in Spin-1/2 Kagome Quantum Spin Liquid Materials}

\author{Aaron T. Breidenbach}
\altaffiliation[]{These authors contributed equally to this work}
\affiliation{Stanford Institute for Materials and Energy Sciences, SLAC National Accelerator Laboratory, Menlo Park, CA 94025, USA}
\affiliation{Department of Physics, Stanford University, Stanford, CA 94305, USA}
\author{Arthur C. Campello}
\altaffiliation[]{These authors contributed equally to this work}
\affiliation{Stanford Institute for Materials and Energy Sciences, SLAC National Accelerator Laboratory, Menlo Park, CA 94025, USA}
\affiliation{Department of Applied Physics, Stanford University, Stanford, CA 94305, USA}
\author{Jiajia Wen}
\author{Hong-Chen Jiang}
\affiliation{Stanford Institute for Materials and Energy Sciences, SLAC National Accelerator Laboratory, Menlo Park, CA 94025, USA}
\author{Daniel M. Pajerowski}
\affiliation{Neutron Scattering Division, Oak Ridge National Laboratory, Oak Ridge TN 37830, USA}
\author{Rebecca W. Smaha}
\affiliation{Stanford Institute for Materials and Energy Sciences, SLAC National Accelerator Laboratory, Menlo Park, CA 94025, USA}
\affiliation{Department of Chemistry, Stanford University, Stanford, CA 94305, USA}
\altaffiliation[Present address: ]{National Renewable Energy Laboratory, Golden, CO 80401, USA}
\author{Young S. Lee}
\email{youngsl@stanford.edu}
\affiliation{Stanford Institute for Materials and Energy Sciences, SLAC National Accelerator Laboratory, Menlo Park, CA 94025, USA}
\affiliation{Department of Applied Physics, Stanford University, Stanford, CA 94305, USA} 

\date{\today}

\begin{abstract}
A quantum spin liquid (QSL) is an exotic quantum state of matter characterized by fluctuating spins which may exhibit long-range entanglement. Among the possible host candidates for a QSL ground state, the $S$=1/2 kagome lattice antiferromagnet is particularly promising. Using high resolution inelastic neutron scattering measurements on  Zn-barlowite (Zn$_\mathrm{x}$Cu$_\mathrm{4-x}$(OD)$_\mathrm{6}$FBr, $x\simeq 0.80$), we measure a spin excitation spectrum consistent with a QSL ground state. Continuum scattering above $\sim$1~meV matches that of herbertsmithite (Zn$_\mathrm{x}$Cu$_\mathrm{4-x}$(OD)$_6$Cl$_2$, $x\simeq 0.85$), another prominent kagome QSL material, indicating universal spinon excitations. A detailed analysis of the spin-spin correlations, compared with density matrix renormalization group calculations, further indicate a QSL ground state for the physically relevant Hamiltonian parameters. The measured spectra in Zn-barlowite are consistent with gapped behavior with a gap size $\Delta = 1.1(2)$~meV. Comparison with a simple pair correlation model allows us to clearly distinguish intrinsic kagome correlations from impurity-induced correlations. Our results clarify the behavior that is universal within this important family of QSL candidate materials.
\end{abstract}

\maketitle

%%%%%%%%%%%%%
% Main Text %
%%%%%%%%%%%%%

A quantum spin liquid (QSL) is a fascinating ground state that may emerge when a collection of quantum magnetic moments are prevented from ordering due to frustration\cite{Broholm2020,Savary2017, Anderson1973}. In such a state, the quantum spins can be correlated without breaking conventional symmetries and may possess long-range quantum entanglement \cite{Broholm2020,Savary2017}. The resulting properties could have far-reaching implications, such as in applications for topological quantum computing \cite{Broholm2020,Kitaev2003}. Improved understanding may also elucidate the fundamental physics in unconventional superconductors \cite{Anderson1987,Imajo2021} as carrier doped QSL's are hypothesized to themselves exhibit high-temperature superconductivity \cite{Anderson1987,Kivelson1987,Jiang2021PRL}. Continued progress requires expanding the family of materials with clear signatures of QSL behavior. 

A promising class of quantum spin liquid candidates exists on two-dimensional frustrated lattices, such as triangular and kagome lattices. The $S$=1/2 kagome mineral herbertsmithite (ZnCu$_3$(OH)$_6$Cl$_2$) is found to avoid magnetic ordering even at temperatures down to 50~mK \cite{Helton2007,Olariu2008}, and inelastic neutron scattering has uncovered fractionalized spinon excitations in the material consistent with QSL behavior \cite{Han2012}. Nuclear magnetic resonance (NMR) and neutron scattering measurements further indicate an $\sim$1~meV energy gap in the QSL state \cite{Fu2015,Han2016,wang2021}, though some NMR results indicate the absence of a gap \cite{Khuntia2020}. Despite controversy surrounding a gap, which include theoretical calculations of possible gapped and gapless QSL ground states in spin-1/2 kagome systems \cite{Jiang2008,Yan2011,Depenbrock2012,Jiang2012, Gong2015,He2017, Liao2017}, the experimental results have established herbertsmithite as a leading candidate for a QSL material \cite{Broholm2020,Norman2016}.

The discoveries in herbertsmithite have motivated intense investigation into similar kagome lattice materials. Zn-substituted barlowite (Zn$_{}$Cu$_{3}$(OH)$_6$FBr) \cite{Han2014, Feng2017, Smaha2020, Tustain2020,  Wang2022, Smaha2023, Campello2025}, like herbertsmithite, is comprised of magnetically isolated kagome layers of antiferromagnetic spin-1/2 Cu$^\mathrm{2+}$ ions. Zn-barlowite differs from herbertsmithite in having a simpler A-A stacking of kagome layers -- compared to herbertsmithite's A-B-C stacking. Both materials have a residual fraction of Cu$^\mathrm{2+}$ impurities on the interlayer sites due to incomplete Zn$^\mathrm{2+}$ substitution, however there is no evidence of impurities within the kagome layers.\cite{Freedman2010,Smaha2020_PRM} Moreover, the Cu$^\mathrm{2+}$ impurities are shifted to lower symmetry positions in Zn-barlowite relative to the impurity positions in herbertsmithite. Susceptibility and NMR experiments on Zn-barlowite show similar properties as herbertsmithite and are suggestive of QSL physics, such as a lack of magnetic ordering down to $T$=100~mK \cite{Smaha2020} and the presence of gapped singlets \cite{wang2021,Wang2022,yuan2022}. Resonant inelastic x-ray scattering on small Zn-barlowite crystals indicated a broad continuum of spin excitations up to $\sim$200~meV\cite{Smaha2023}.

In this article, we present high-resolution inelastic neutron scattering measurements on large Zn-barlowite crystals. We discover a continuum of excitations for the kagome magnetic moments, indicative of fractionalized spinon excitations. This matches the same signature found in herbertsmithite, and therefore points to universal physics for these $S=\frac{1}{2}$ kagome antiferromagnets. The measured intensities provide unprecedented insight into the relevant spin-spin correlations, which allows us to clearly distinguish intrinsic kagome correlations from impurity-induced correlations. Comparisons with numerical calculations further support the case for a QSL ground state. Finally, the data are consistent with the presence of a spin gap of $\sim J/15$, suggesting that a gapped quantum spin liquid is present in these materials.

%%%%%%%%%%%%%%%%%%%%%%
% NEUTRON SCATTERING %
%%%%%%%%%%%%%%%%%%%%%%

\section*{Spin Excitations Measured by Inelastic Neutron Scattering}

% order:
% -- high E
% -- low 
% -- model
% section in figure caption

Using a new crystal synthesis technique, we grew large (up to $\sim$5$\times$5$\times$0.1~mm), high quality single crystals of deuterated Zn-barlowite (Zn$_\mathrm{x}$Cu$_\mathrm{4-x}$(OD)$_\mathrm{6}$FBr, $x\simeq 0.80$). This development enabled our inelastic neutron scattering measurements on a large 0.76~g co-aligned array of $\sim$190 of these crystals (Figure \ref{fig:figE1} in extended data). Figure 1 shows representative magnetic scattering intensities in the (HK0) zone for Zn-barlowite, compared to that of herbertsmithite, at $T$=1.7~K. The Zn-barlowite signal, proportional to the dynamic structure factor $S(\mathbf{q},\omega)$, is shown in the left halves of panels (a) and (c) at energy transfers of $\hbar\omega$=0.4~meV and $\hbar\omega$=1.3~meV, respectively (after background subtraction and symmetrization detailed in Methods and figure \ref{fig:dproc} in extended data). The left halves of panels (b) and (d) show similar measurements on herbertsmithite using data from Han et al. \cite{Han2016}.

\begin{figure}[b]
\includegraphics[width=0.47\textwidth]{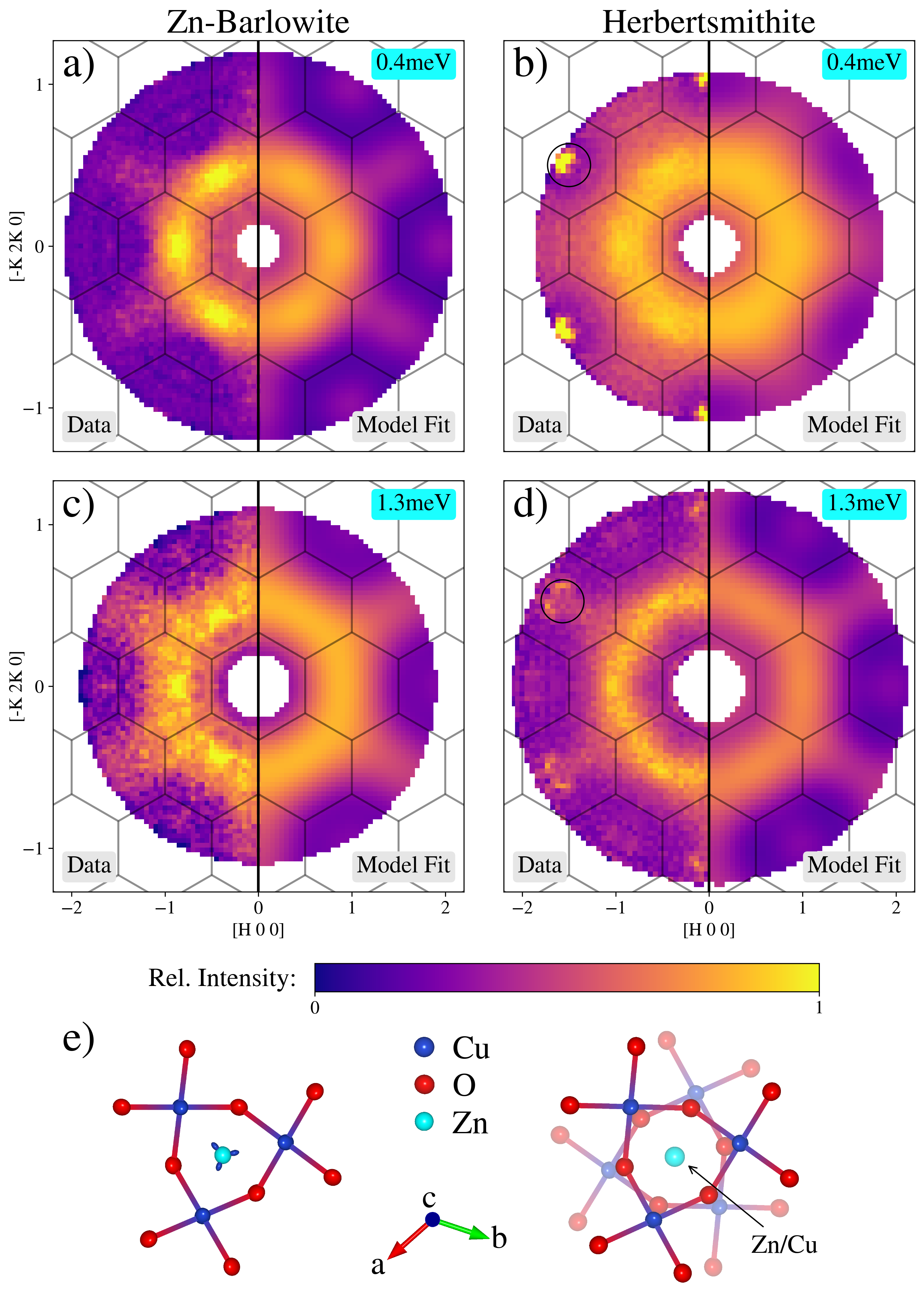}
\caption{\label{fig:fig1} \textbf{Comparison of magnetic excitations between candidate kagome QSL materials, showing universal behavior, with distinct low-energy impurity-related scattering.}
\textbf{(a-d)} Comparisons between measured and modeled magnetic neutron scattering intensity for Zn-barlowite (a, c) and herbertsmithite (b, d) \cite{Han2016} for energy transfers of  $\hbar\omega$=0.4~meV and $\hbar\omega$=1.3~meV, respectively (see the Spin Correlations and Numerical Comparisons section). Black circles in b and d indicate scattering from phonons or tails of the Bragg peak intensity near the [1 1 0] positions, which are not captured by the magnetic model. All data shown were taken with incident energy $E_i$=3.32~meV and $T$=1.7~K. Zn-barlowite data was taken at the Cold Neutron Chopper Spectrometer at Oak Ridge National Laboratory and contours in (a, c) are respectively integrated over [0.3,0.5]~meV and [1.2,1.4]~meV to match the resolution of the data from Han et al. \cite{Han2016}. \textbf{(e)} Crystal structures of two triangular plaquette layers of Zn-barlowite (left) and herbertsmithite (right). Cu, O, and Zn atoms are shown to highlight differing interlayer impurity positions and kagome layer stacking between the materials.}
\end{figure}

The higher energy 1.3~meV scattering reveal patterns that are very similar between Zn-barlowite and herbertsmithite, suggesting universal behavior. In contrast, the lower energy 0.4~meV scattering patterns differ notably, such as the intensity variations along the (1,0,0) versus the (1,1,0) directions. Panel (e) depicts the crystal structure of the materials, highlighting the differences between the interlayer impurity positions. In contrast to the centered impurities in herbertsmithite, the impurities in Zn-barlowite are shifted to lower symmetry positions and are expected to couple to the nearest kagome moments differently \cite{Smaha2020, Jeschke2015}. Our results suggest that the low energy scattering originates primarily from the impurity-kagome correlations which are different between the materials \cite{Han2016, Smaha2020}, while the higher energy scattering originates primarily from the kagome planes, which are structurally nearly identical. These observations indicate that the intrinsic kagome moments in both materials exhibit similar universal QSL excitations.

Modeling the observed scattering patterns allows us to make a quantitative identification of the intrinsic versus the impurity-induced spin correlations \cite{Han2016}. The right halves of panels (a-d) in Figure \ref{fig:fig1} model the corresponding left-half data using a weighted sum of contributions from only the six bond types with the shortest in-plane distances. This model confirms that the high energy scattering is dominated by kagome layer correlations, while the low energy scattering in Zn-barlowite is dominated by impurity to kagome correlations. Further details on the spin correlations and the empirical model are discussed below.

\begin{figure}[b]
\includegraphics[width=0.47\textwidth]{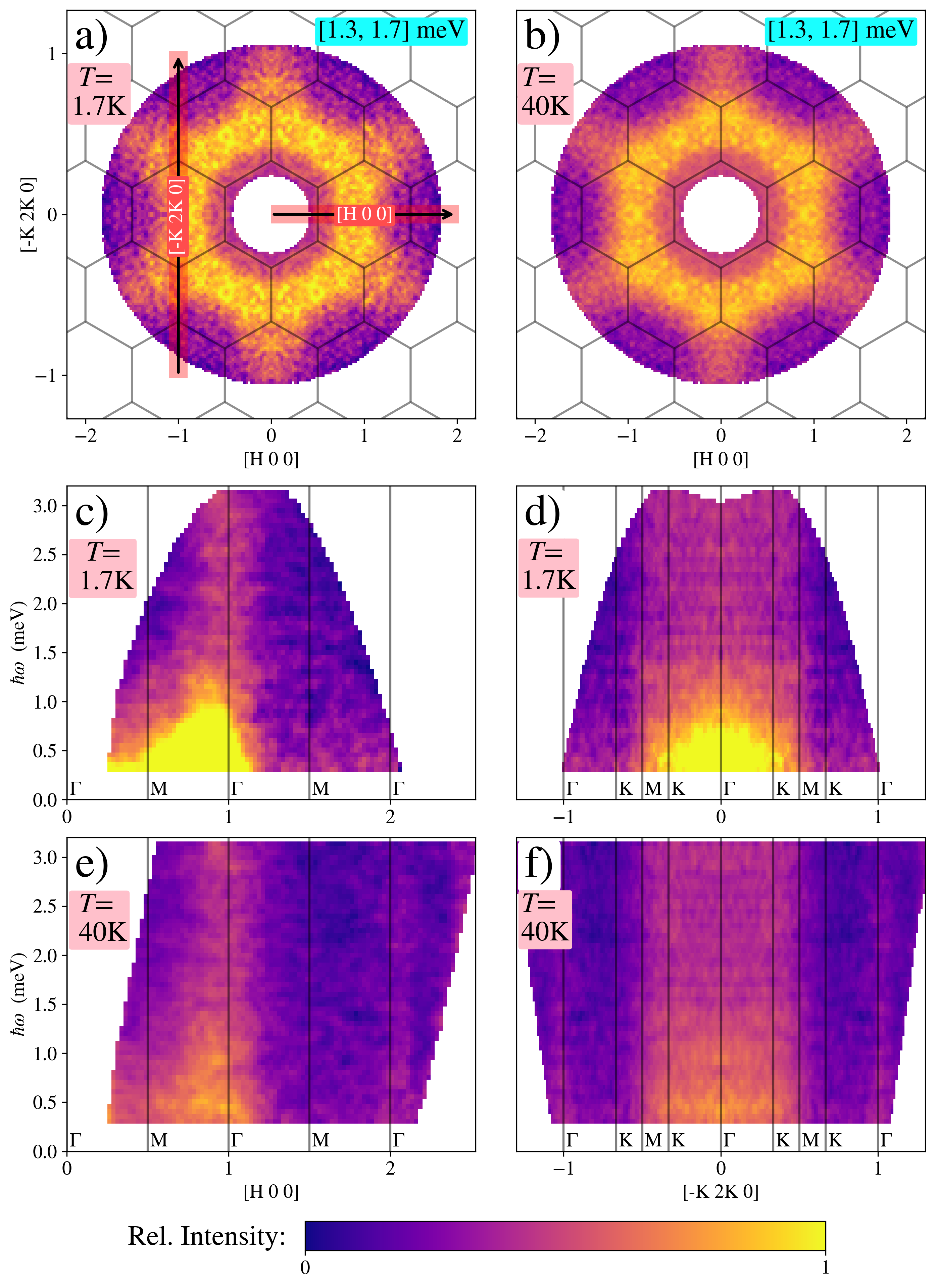}
\caption{\label{fig:fig2} \textbf{Spinon continuum in Zn-barlowite, highlighting its relative independence on energy transfer and robustness to changes in temperature.} \textbf{(a, b)} Inelastic scattering pattern integrated over $\hbar\omega$=[1.3,~1.7]~meV at $T$=1.7~K (a) and $T$=40~K (b). \textbf{(c-f)} Momentum-integrated scattering over regions indicated by red boxes in (a) at $T$=1.7~K (c, d) and $T$=40~K (e, f). Data shown were all taken with incident energy $E_i$=3.32~meV. (a, b) share a common intensity scale as do (c-f).}
\end{figure}

A more detailed look at the dynamic structure factor $S(\mathbf{q},\omega)$ in shown in Figure \ref{fig:fig2}. Panels (a,b) show the intensity integrated over the range $\hbar\omega$=[1.3,~1.7]~meV at $T$=1.7~K and $T$=40~K. Notably, the patterns appear similar to each other and to that in Figure \ref{fig:fig1}(c), demonstrating that the elevated temperature of $T\sim J/5$ minimally affects the spin excitations. Additional data at $T$=1.7~K and $T$=40~K appear in Figure \ref{fig:figExpo} in the Extended Data. Panels (c-f) depict cuts through the spectrum along various symmetry directions in reciprocal space. Here, clear continuum scattering is observed, consistent with fractionalized spinon excitations \cite{Han2012}. Temperature-dependent changes between the data only appear below $\sim$1.2~meV where impurities start to play a role, highlighting the temperature robustness of the intrinsic kagome scattering at higher energies. These data also show that the spinon excitations in Zn-barlowite (from $\sim1.2$ to $\sim3$ meV) have remarkably little dependence on energy, similar to observations in herbertsmithite \cite{Han2012}.

\begin{figure}[h]
\includegraphics[width=0.47\textwidth]{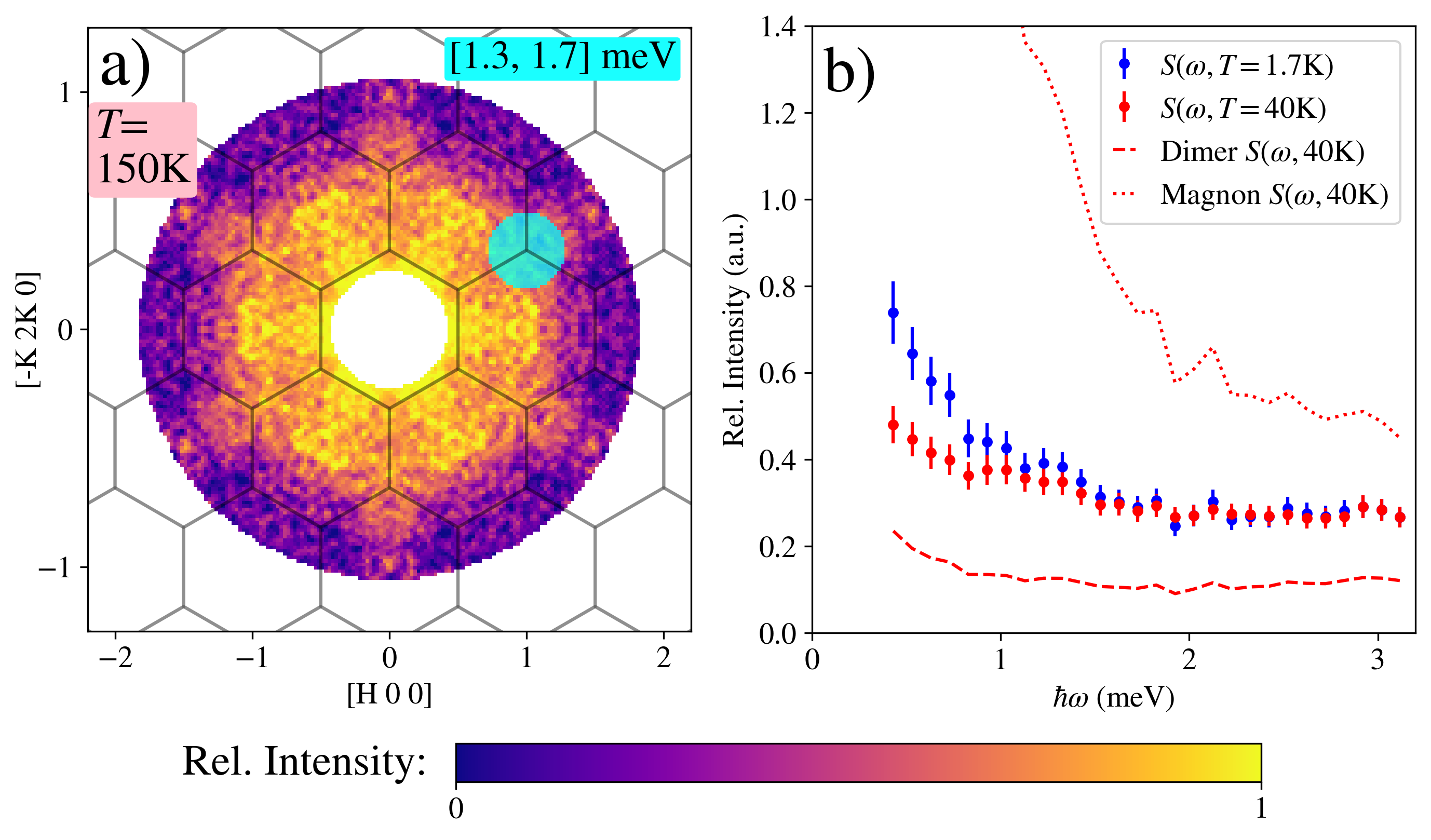}
\caption{\label{fig:fig3} \textbf{Temperature dependence of spinon excitations.} \textbf{(a)} Inelastic scattering pattern integrated over $\hbar\omega$=[1.3,~1.7]~meV at $T$=150~K showing persistent star-shaped scattering. \textbf{(b)} Energy dependent magnetic scattering intensity integrated over the cyan-highlighted $\mathbf{q}$-region in (a) for sample temperatures $T$=1.7~K and $T$=40~K. Error bars for $T$=1.7~K and $T$=40~K are statistical counting errors. Line plots show estimated scattering for $T$=40~K based on the $T$=1.7~K data scaled by statistics for different models: (dashed) a dimer model with a singlet ground state and triplet excited state; and (dotted) those associated with magnons. Both are shown to diverge from the observed data.}
\end{figure}

At an even higher temperature, Figure \ref{fig:fig3}(a) shows the scattered intensity measured at $T$=150~K integrated over the energy range $\hbar\omega$=[1.3,~1.7]~meV. The pattern is similar to the $T=1.7$~K data in Figure \ref{fig:fig1}(c), implying intrinsic kagome correlations persist up to temperatures approaching $J$. Panel (b) shows the energy dependence of $S(\mathbf{q},\omega)$ integrated over the $\mathbf{q}$-region highlighted in panel (a) for temperatures $T$=1.7~K and $T$=40~K. The overlap of the data above $\sim$1.5~meV suggests the scattering does not follow the thermal statistics of conventional magnetic systems. The dotted line shows the expected $T$=40~K scattering derived from applying the thermal factor associated with magnons to the $T$=1.7~K data. The clear disagreement is consistent with the lack of magnetic ordering in the system. Alternatively, for a system with a ground state and an excited state that are both highly degenerate, the ratio of intensities would follow $S(\omega, T_1)/S(\omega, T_2)=\left[n_d+\exp(-\beta_2\hbar\omega)\right]/\left[n_d+\exp(-\beta_1\hbar\omega)\right]$, where $n_d$ gives the ratio of the ground state degeneracy to the excited state degeneracy. An isolated dimer model with a triplet excited state and singlet ground state would have $n_d=1/3$; the solid line in Figure \ref{fig:fig3}(b) shows the expected $T$=40~K scattering for this model which also fails to describe the data. Fitting the data above 1.5~meV (to avoid the effects of impurities), we find that a large relative ground state degeneracy would be required ($n_d\ge30$) to best describe the measured data. Hence, neither the thermal behavior of magnons nor that of isolated dimers can describe the observed spin excitations.

%%%%%%%%%%%%%%%%%%%%%
% SPIN CORRELATIONS %
%%%%%%%%%%%%%%%%%%%%%

\section*{Spin Correlations and Numerical Comparisons}

To derive more quantitative information about the spin correlations, we employ an empirical model that allows us to extract correlations between kagome moments separately from correlations involving impurity moments. Our approach assumes
\begin{eqnarray} 
S_{\mathrm{mag}}(\mathbf{q}, \omega)\approx \alpha(\omega) |F(\mathbf{q})|^2\left(1+2\sum^n_{i=1}\rho_i(\omega) f_i(\mathbf{q})\right),
\label{eq:1}
\end{eqnarray}
where $|F(\mathbf{q})|^2$ is the square magnetic form factor, and $f_i(\mathbf{q})$ is defined by a sum over equal-length bonds with label $i$ between Cu$^{2+}$ ions that exists in $d_i$ distinct directions $\{\mathbf{r}_{i1},\dots, \mathbf{r}_{id_i}\}$: $f_i(\mathbf{q})\!\!=\!\!(1/d_i)\sum^{d_i}_{n=1}\cos(\mathbf{q}\!\cdot\!\mathbf{r}_{in})$. This means we assume a Fourier transform encodes the momentum-dependent scattering at a given energy. $\alpha(\omega)$ and all $\rho_i(\omega)$'s for included bonds are parameters estimated from fits to the data. $\alpha(\omega)$ gives the energy dependence of the overall scattering intensity and estimates the momentum-integrated dynamic structure factor $S(\omega)$, since the integral of $f_i(\mathbf{q})$ over all $\mathbf{q}$ is zero for any bond. Parameters $\rho_i$ for each bond are proportional both to the bond density and the average spin-spin correlation over the bond -- see Supplementary Information for further details. The bond density's energy independence means $\rho_i(\omega)$ is proportional to the energy-dependent spin-spin correlation, i.e. $\rho_i\propto |\left\langle SS'\right\rangle_i|$. We estimate $\alpha$ and $\rho_i$ from our data at each energy, with standard errors, using weighted least squares linear regressions; see Methods for details.

\begin{figure}[t]
\includegraphics[width=0.47\textwidth]{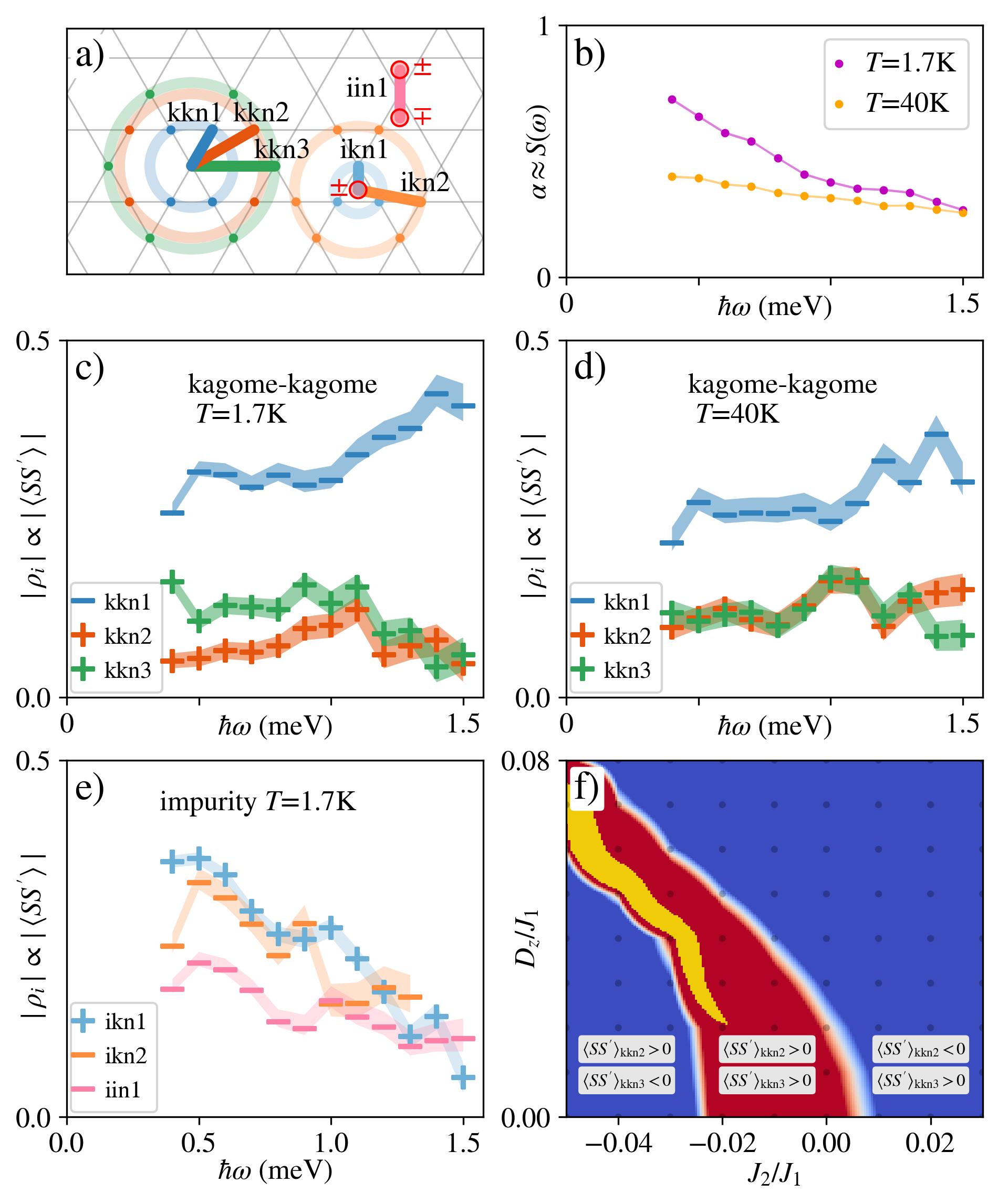}
\caption{\label{fig:fig4} \textbf{Extracted pair-wise spin correlations over energy and temperature.} \textbf{(a)} Illustration of bond directions over which we report estimated correlation signals. Red circles indicate impurities with $\pm$ indicating their c-axis displacements from the kagome plane. \textbf{(b)} Fitted parameter $\alpha\left(\omega\right)\approx S(\omega)$ for $T$=1.7~K and $T$=40~K with standard errors. (c, d) Fitted relative correlation parameter $\rho(\omega)$ for three nearest kagome-kagome bonds for $T$=1.7~K and $T$=40~K. Here, AFM correlations are denoted with a ``$-$'' symbol and FM with a ``$+$'' symbol. (e) Fitted $\rho(\omega)$ for three nearest impurity-involved bonds for $T$=1.7~K data. (f) DMRG results for second and third nearest-neighbor correlations. Red region indicates parameter space where both are positive, consistent with fitted $\rho(\omega)$ signs. Yellow region indicates parameter combinations that most closely match $\hbar\omega$=[1.3,~1.7]~meV neutron data as described in the text. Reported errors in (b-e) reflect standard errors from parameter fits. See Methods and Supplementary Information for details.}
\end{figure}

Figure \ref{fig:fig4} shows results from fitting Equation \ref{eq:1} to the magnetic scattering data. Panel (a) illustrates kagome plane projections of the six bonds used in the fitting, which are the shortest six in Zn-barlowite. These include the first three kagome-plane nearest-neighbors -- labeled ``kkn1-3'' -- the shortest two impurity-to-kagome bonds, ``ikn1-2,'' and the bond between neighboring impurities, ``iin1.'' The models in Figure \ref{fig:fig1} are made from equation \ref{eq:1} including these bonds and their isolated $f_i(\mathbf{q})$ contributions are shown in Figure \ref{fig:qmaps} in Extended Data. Panel (b) of Figure \ref{fig:fig4} shows the fitted $\alpha(\omega)\approx S(\omega)$ at low and elevated temperatures.

A highlight of the data analysis is the ability to isolate and extract the kagome-kagome spin correlations. Figure 4(c, d) report fitted $|\rho_i|$ correlation parameters for the three shortest intrinsic kagome bond correlations for $T$=1.7~K and $T$=40~K. Over both temperatures and all energies, these fitted parameters indicate strong antiferromagnetic (AFM) spin correlations over nearest-neighbor kagome bonds, consistent with the known dominant nearest-neighbor AFM interactions. Moreover, we extract reliable correlations over both second and third nearest-neighbors which are weaker and ferromagnetic (FM). This combination of further neighbor correlations differ from those expected in $q$=0 and $\sqrt{3}\times\sqrt{3}$ ordered states in AFM kagome magnets \cite{Grohol2005}. The nearest-neighbor kagome correlations stay relatively flat with increasing energy before rising above $\sim$1~meV, concomitant with the further neighbor correlations weakly decreasing. The energy dependence is more subdued in the elevated $T$=40~K dataset. 

Figure 4(e) plots the fitted $\rho_i$ correlation parameters for impurity-involved correlations at $T$=1.7~K. The impurity correlations decrease almost monotonically with energy and are only significant below $\sim1.5$ meV, consistent with the weakly coupled nature of the interlayer impurties \cite{Han2012,Han2016}. At $T=40$~K (not shown), only the nearest impurity-kagome correlation is significant and also decreases monotonically. Exact coefficient comparisons with herbertsmithite are reported in \ref{tab:compths} in Extended Data.

The quality of the extracted correlations allows for close comparison with DMRG calculations based on a model $S=1/2$ Hamiltonian for the material. For the $\rho_i$ correlation parameters corresponding to bonds in the kagome plane, we can make direct comparisons to DMRG calculated spin-spin correlations after weighting by the bond abundances. For these calculations, we parameterize the spin Hamiltonian \cite{Han2012PRL}
\begin{eqnarray}\label{Eq:Ham}
H = J_1 \sum_{\left\langle ij\right\rangle}\vec{S}_i\!\cdot\!\vec{S}_j + J_2\!\sum_{\left\langle\left\langle ij\right\rangle\right\rangle}\! \vec{S}_i\!\cdot\!\vec{S}_j + \sum_{\langle ij\rangle}\vec{D}_{ij}\!\cdot\!(\vec{S}_i\!\times\!\vec{S}_j ), 
\end{eqnarray}
with $J_1$ and $J_2$ representing couplings over nearest-neighbor and second-nearest-neighbor kagome sites and $D_z$, the Dzyaloshinskii–Moriya (DM) interaction with the DM vector pointing out-of-plane \cite{Lee2013}. Setting $J_1>0$, panel (f) of Figure \ref{fig:fig4} shows a contour map over unitless quantities $J_2/J_1$ and $D_z/J_1$. The red region indicates parameter combinations where DMRG yields positive second and third nearest-neighbor kagome spin correlations, as observed in the data. This comparison suggests that $J_2$ is ferromagnetic with magnitude less than $\sim$5\% of $J_1$, assuming a 0.1$J_1$ bound on $D_z$. 

For a more quantitative comparison, we integrate the $T$=1.7~K data over $\hbar\omega$=[1.3,~1.7]~meV (a range with minimal impurity influence) to determine the kagome-only correlations with improved statistics. We then use the calculated DMRG nearest-neighbor correlations to set the absolute scale and let the sum of squared differences in the other two correlations be a loss metric. Setting a conservative threshold yields the yellow region in Figure \ref{fig:fig4} representing parameters that most closely align with the scattering data. This tighter region again indicates that $D_z$ and $J_2$ are nonzero but small compared to $J_1$. The case $J_2/J_1\gg 0$ would drive $q=0$ ordering and the opposite case $J_2/J_1\ll 0$ would drive $\sqrt{3}\times\sqrt{3}$ ordering over the kagome lattice \cite{Suttner2014}. The parameter space for $J_2$ and $D_z$ that we deduce are consistent with Zn-barlowite being in an intermediate region of the phase diagram away from either type of order, consistent with a quantum spin liquid.

\begin{figure}[b]
\includegraphics[width=0.47\textwidth]{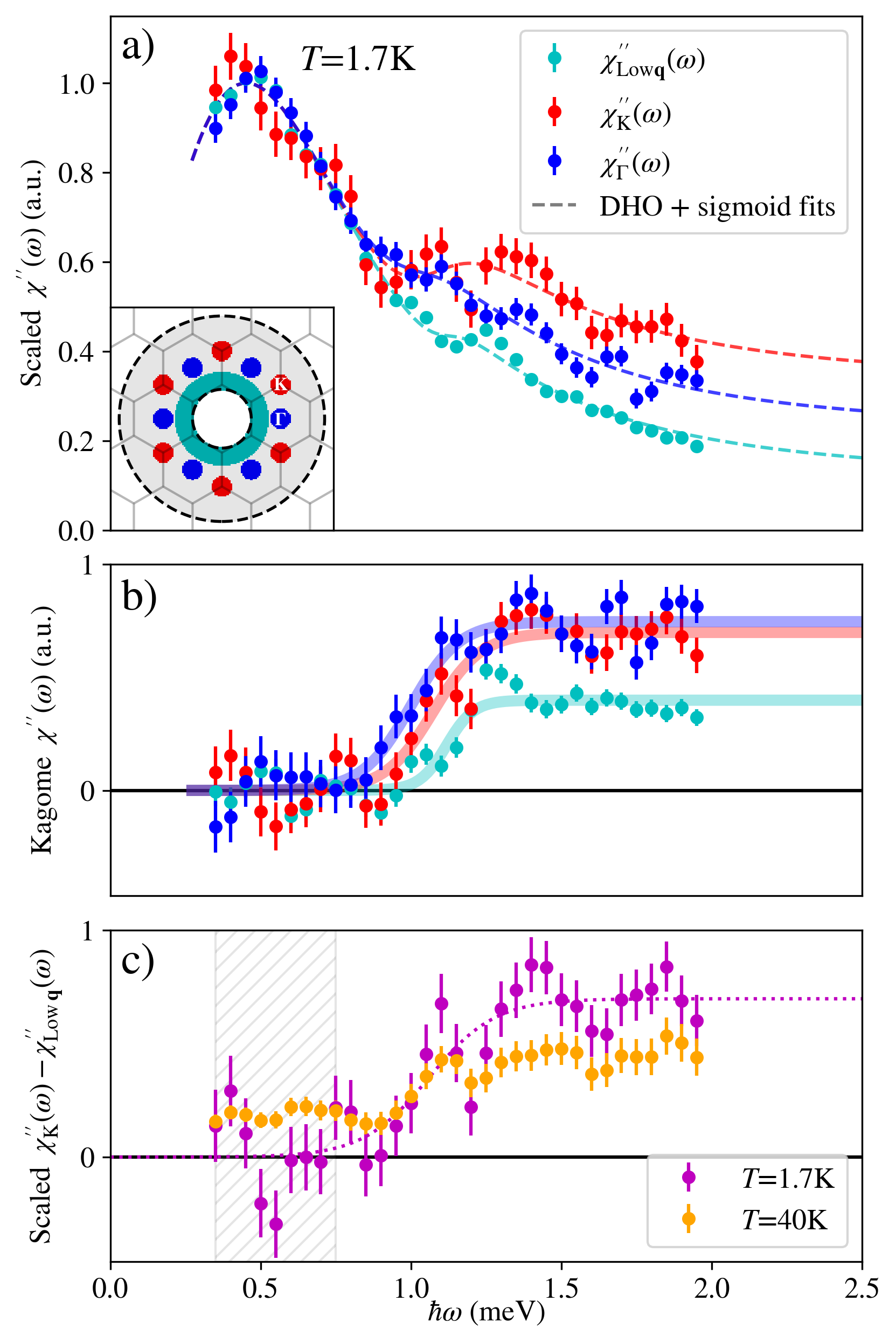}
\caption{\label{fig:fig5} \textbf{Signature of a possible spin gap in intrinsic kagome spin excitations.} \textbf{(a)} Energy cuts of $\chi''(\omega)$ over various high-symmetry positions scaled and modeled as described in the text. Fitted function curves for each are shown as dashed lines. Inset shows integration regions inscribed by the $q$-region kinematically accessible at all reported energies. \textbf{(b)} non-normalized data from (a) with DHO component subtracted. Resulting sigmoid fits are highlighted showing similar widths and centers consistent with a $1.1(2)$~meV energy gap. \textbf{(c)} Subtractions of $\chi''_{\mathrm{K}}(\omega)$ from $\chi''_{\mathrm{Low}-\mathbf{q}}(\omega)$, where the latter is scaled such that the mean of the $T$=1.7~K data subtraction is zero at energies in the hatched region. Subtractions with this same scale are reported for $T$=1.7~K and $T$=40~K data along with a fitted sigmoid for the former. Error bars in (a-c) are propagated from counting statistics.}
\end{figure}

%%%%%%%%%%%%%%%%
% GAP ANALYSIS %
%%%%%%%%%%%%%%%%

\section*{Kagome Spin Gap Analysis}

In herbertsmithite, after impurity-induced effects are taken into account, the intrinsic QSL state is believed to have a spin gap characterized by an energy $\sim$0.8~meV based on neutron scattering \cite{Han2016} and NMR \cite{Fu2015} measurements, with indications of spatial inhomogeneity of the gap size \cite{wang2021}. Determining whether the QSL state in Zn-barlowite is similarly gapped will also require accounting for the impurity-induced correlations. While the analysis shown in panels (c, d) of Figure \ref{fig:fig4} already hints at changes in the kagome-kagome spin correlations around $\sim$1~meV, the persistance of the impurity-kagome correlations prevent a direct observation of kagome spin gap behavior.

To determine whether a spin gap exists in Zn-barlowite, we employed the fluctuation dissipation theorem to derive $\chi''(\mathbf{q}, \omega)$ from $S(\mathbf{q},\omega)$. Following the approach of Han et al. \cite{Han2016} we assume that $\chi''(\mathbf{q}, \omega)$ is the sum of contributions from impurity and intrinsic kagome contributions that are separable in $\mathbf{q}$ and $\omega$. Next, we model the energy dependence of the impurity contribution as a damped harmonic oscillators (DHO) and that of the kagome contribution as a sigmoid function (see Methods for details). 

Panel (a) of Figure \ref{fig:fig5} shows $\chi''(\omega)$ from $T$=1.7~K data integrated over various high-symmetry $\mathbf{q}$-regions shown in its inset; all are scaled such that the maximum of the DHO component is 1. The panel shows that below $\sim$0.8~meV, the scaled $\chi''$ curves behave similarly and adhere well to a common DHO which captures the energy-dependence of the impurity scattering. Panel (b) shows scattering data with the DHO component subtracted and without scale normalization. The independently fit sigmoid components all center at similar energies $1.1(2)$~meV and have similar widths.

From our correlation analysis in the previous section, we know that nearest impurity-to-kagome bonds contribute significantly to the low-$\mathbf{q}$ scattering and that the nearest-neighbor kagome spin correlations yield intensity at the depicted K-regions (also see Figure \ref{fig:qmaps} in Extended Data). This is consistent with the intensities of the DHO function capturing the impurity scattering and the sigmoid function capturing the kagome correlations. The results shown in panels (a, b) of Figure \ref{fig:fig5} are hence indicative of a gapped QSL state in Zn-barlowite with gap energy $\sim$1.1~meV. Figure \ref{fig:chippGap} in Extended Data further shows this model applied to the derived $\mathbf{q}$-integrated $\chi''(\omega)$ with higher statistics, again showing a $\sim$1.1~meV energy gap.

Figure 5(c) shows a similar analysis without any explicit assumption of a DHO impurity fit. Using the $T$=1.7~K data, the $\chi''(\omega)$ corresponding to low-$\mathbf{q}$ is scaled to match $\chi''_K(\omega)$ in the energy range $\hbar\omega=[0.35,~0.75]$~meV and subtracted from $\chi''_K(\omega)$. This scaled difference (magenta) follows a sigmoid with a similar $\sim$1.1~meV center to those in panel (b). The same process is applied to the $T$=40~K data and shown in orange. Interestingly, the high temperature data shows diminishment of the scattering intensity below $\sim$1.1~meV, however the intensity does not fall to zero. While this behavior indicates the system is not gapped at $T$=40~K, consistent with NMR results that show the gapped singlet fraction only grows below $T$=40~K \cite{Wang2022}, the reduction in scattering intensity suggests the persistence of a pseudo-gap-like energy scale on the order of the spin gap energy at elevated temperatures.

\section*{Conclusion}
Our inelastic neutron scattering data affirm that Zn-barlowite exhibits signature features of a QSL, and comparison with results on herbertsmithite allows us to identify universal behavior common to these leading kagome QSL materials. The spin excitations of the kagome moments, at energies above the impurity-dominated regimes, are strikingly similar and take the form of a spinon continuum. Our pair correlation model allows us to clearly distinguish intrinsic kagome correlations from impurity-induced correlations. In Zn-barlowite, the intensity pattern is dominated by relatively short-range kagome correlations, shows little change with energy (at least for energies below $J_1/5$), and persists to at least $T$=150~K. Our results also point to a spin gap with $\Delta = 1.1(2)$~meV, which, along with similar evidence in herbertsmithite, \cite{Han2016,Fu2015,wang2021} lends credence to a gapped ground state for this family of S=1/2 kagome QSL materials. While the intensity of the dynamic structure factor changes appreciably with $\mathbf{q}$, the measured energy gaps and gap widths are relatively independent of $\mathbf{q}$. Future measurements to uncover finer-grained details on the distribution of gap features over $\mathbf{q}$ would aid our understanding of the gap inhomogeneity deduced in recent NMR measurements on these materials \cite{wang2021}.

The measured first, second, and third nearest-neighbor kagome-kagome correlations place Zn-barlowite within the QSL phase of the extended phase diagram (involving $J_2$ and $D_z$) calculated by DMRG. This implies constraints on the interactions where $|J_2|$ and $|D_z|$ are both small ($< J_1/10$), indicating the Hamiltonian is well described by the nearest-neighbor Heisenberg model to first approximation. This is consistent with other measurements refining the Hamiltonian of herbertsmithite \cite{Han2012PRL, Zorko2008}. Although the interactions are small, the combination of negative $J_2$ and positive $D_z$ serves to stabilize the QSL phase. This information sheds light on comparisons with theoretical dynamic structure factor results. DMRG \cite{Zhu2019} and variational Monte Carlo \cite{Ferrari2023} calculations on the kagome AFM system assuming $J_2=D_z=0$ predict neutron intensities that show some quantitative differences with our data, such as higher intensity near the $\Gamma$ position relative to our results in Figure 5. Including the effects of negative $J_2$ and positive $D_z$ would likely flatten this intensity variation to more closely match the data. A small easy-axis exchange anisotropy similar to that in herbertsmithite \cite{Han2012PRL} may also be needed. Theoretical work which includes coupling between spinons and vison excitations has also been proposed to explain excitations in herbertsmithite \cite{Punk2014} and appears to match our results. Adjustment of the phenomenological parameters in this model could lead to even closer alignment.

A virtue of both herbertsmithite and Zn-barlowite is that the spin Hamiltonians are relatively well characterized \cite{Han2012PRL,Smaha2020} with the effects of the interlayer impurities having been well studied \cite{Han2016,wang2021,yuan2022}. As a result of our finding that both have a similar spinon continuum, it is important to investigate whether this universality pertains to other candidate kagome QSL materials such as Zn-doped claringbullite (ZnCu$_3$(OH)$_3$FCl) \cite{Feng2018, Georgopoulou2023}. Identifying behavior that is truly universal would also help us understand the differences with recent results within the kagome family that includes YCu$_9$(OH)$_{19}$Cl$_8$ \cite{Chatterjee2023} and YCu$_3$(OH)$_6$Br$_2$[Br$_{1-x}$(OH)$_x$]  \cite{Chen2020, Liu2022, Zeng2024}. While YCu$_9$(OH)$_{19}$Cl$_8$ orders around 2.1 K, YCu$_3$(OH)$_6$Br$_2$[Br$_{1-x}$(OH)$_x$] does not order down to 50 mK and has been studied as another QSL candidate with an entirely different type of structural disorder. Both materials show dispersive spin excitations measured by inelastic neutron scattering emanating from $Q=(\frac{1}{3},\frac{1}{3})$-type positions. Improved understanding of the spin Hamiltonians of these two materials would aid in the classification of universal QSL behaviors. Our results show that Zn-barlowite and herbertsmithite have robust and consistent QSL physics, and they represent a most promising class of kagome QSL materials. 

\section*{Acknowledgements}
We thank S. Kivelson and W. He for insightful discussions. We thank R. Matheu for assistance in collecting crystallography data. We thank D. Burns for assistance with electron microprobe measurements. This work is supported by the U.S. Department of Energy (DOE), Office of Science, Basic Energy Sciences, Materials Sciences and Engineering Division, under contract DE-AC02-76SF00515. This research used resources at the Spallation Neutron Source, Department of Energy (DOE), Office of Science User Facilities, operated by the Oak Ridge National Laboratory. Part of this work was performed at the Stanford Nano Shared Facilities (SNSF), supported by the National Science Foundation under award ECCS-2026822. R.W.S. was supported by a NSF Graduate Research Fellowship (DGE-1656518).

\section*{Competing Interests}
The authors declare no competing interests.

\section*{Author Contributions}
ATB, JW, RWS, and YSL conceived the project. ATB, ACC, JW, and YSL coordinated the project. ATB and RWS synthesized and characterized the Zn-barlowite single crystals. ATB performed the sample assembly with significant help from ACC and JW. ATB, JW, and DMP performed the neutron measurements with help from ACC and YSL. ATB and ACC performed the background subtractions and processing. ACC and JW performed the herbertsmithite quantitative analysis. ATB, ACC, and JW performed the analysis of temperature dependence. ACC and ATB conceptualized and implemented spin correlation models to interpret the data. ACC implemented the linear regression model used to extract the spin correlations. HCJ performed the DMRG calculations. ACC performed the kagome spin gap analysis. ACC led visualization efforts and made final versions of the figures. ACC, ATB, JW, HCJ, and YSL wrote the paper with input from all co-authors.

\bibliography{sample}
\newpage
%\afterpage{\blankpage}

%%%%%%%%%%%
% Methods %
%%%%%%%%%%%

\section*{Methods}
\subsection{Crystal Grown and Sample Preparation}

In the first step of crystal growth, hydrothermal reactions were performed in 45 ml PTFE-lined stainless steel autoclaves. Zn-substituted barlowite powder was synthesized in this reaction using CuO (Aldrich), NH$_4$F (Alfa, 96$\%$), ZnBr$_2$ (BTC, 99.999$\%$), and 20 ml D$_2$O (Aldrich, 99.9$\%$). These contents are heated over 2.5 hrs from 35 $^{\circ}$C to 185$^{\circ}$C, held for 48 hrs, then cooled back to 30$^{\circ}$C over 30h. The complete products of this pre-reaction (including both the Zn-barlowite powder and ion-rich D$_2$O) were placed into a 11.8mm OD thin-walled PFA Teflon liner and frozen. The end of the liner was then vacuum sealed ($\sim$10$^{-2}$ Torr) and the liner and the contents were placed into a quartz tube with an inner diameter of 12.7mm and outer diameter of 18.7mm. With the contents frozen, vacuum was pulled in the quartz tube ($\sim$10$^{-3}$ Torr) and it was sealed. The PFA Teflon liner is necessary for this growth to ensure the present fluorine ions do not react with the quartz tube; this removes them from the reaction and weakens the quartz tube, leading to an explosion hazard. Note that the pre-reaction to form Zn-barlowite powder can also be performed directly in the Teflon-lined test tube; both methods are effective, but the separate pre-reaction in the autoclave seemed to work more consistently for us.

\begin{figure*}[h]
\includegraphics[width=0.5\textwidth]{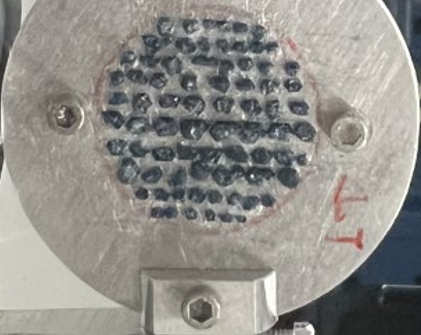}
\caption{\label{fig:figE1} Co-alligned crystals of Zn-barlowite mounted on Aluminum plates that are affixed to the CNCS instrument sample holder. This sample contains 0.76g of Zn-barlowite crystals mounted across three vertically stacked plates. Neutrons were scattered off this crystal array to obtain the data reported, with the rough beam width indicated by the red circle on the plate shown. }
\end{figure*}

Two separate quartz tube assemblies were made with this above process. Each was placed inside a three zone furnace for continued hydrothermal synthesis. The zones formed a temperature gradient with a $\sim$180$^{\circ}$C hot end and a $\sim$170$^{\circ}$C at the cold. The pre-reacted powder is manipulated to start at the hot end. In these furnace conditions, Zn-substituted barlowite powder will dissolve in the D$_2$O fluid at the hot end and will nucleate at the cold end as it is slowly thermally transported there. The growths were monitored regularly and ran for approximately one year. Each growth yielded several clumps of large crystals that were strongly attached to each other at domain walls. Manually separating these clumps yielded single-domain crystals up to $\sim$5$\times$5$\times$0.1mm in size. ICP-AES measurements on the resulting groups of crystals revealed Zn concentrations of $x$=0.78 for batch one and $x$=0.85 for batch 2. Electron microprobe analysis was also performed and corroborated the above measurements, revealing Zn-concentrations $x$=0.75 for batch one and $x$=0.85 for batch 2. 0.57~g of crystals from the first batch and 0.19~g from the second were coaligned on two mounting plates using CYTOP adhesive. This yielded a total sample mass of 0.76~g. Figure \ref{fig:figE1} in the Extended Data shows the final sample assembly. 

\subsection{CNCS Experiment and Data Processing}
The reported neutron data were collected using the CNCS spectrometer at Oak Ridge National Laboratory’s Spallation Neutron Source. Spectra were measured at sample temperatures $T$=1.7~K, $T$=40~K, and $T$=150~K with incident neutron energy $E_i$=3.32~meV over 360$^\circ$ scans in the HK0 scattering plane. The chopper frequency used was 180Hz. 

Two separate background measurements were performed; one with just an empty can of Helium and another on dummy aluminimum sample plates coated with the same mass CYTOP as that of the true sample holder. These two background datasets were subtracted from the crystal data using an empirically derived self-shielding factor. This factor governs only the relative subtraction of the CYTOP signature to the helium and was determined empirically via a method outlined in the Supplementary Information.

Following this step, we used a background subtraction technique introduced by Helton et al. \cite{Helton2007}. This approach utilizes the oddness of $\chi''(\omega)$ with respect to $\omega$ and the fact that $S(\mathbf{q},\omega)=\left[1+n(\omega)\right]\chi''(\mathbf{q},\omega)$, where $n(\omega)$ is the Bose occupation factor. Since the time of flight based CNCS spectrometer measures scattering events that both transfer energy to and from the crystals (positive and negative energy scattering events respectively), we can exploit this oddness to clean the data of all temperature-independent background features. With such data taken at two temperatures $T_{\mathrm{L}}$ and $T_{\mathrm{H}}$ the method can be described by the equation set
\begin{equation}\label{eq:Helton}
\begin{split}
S(\mathbf{q},+\omega,T_{\mathrm{L}})=&S(\mathbf{q},+\omega,T_{\mathrm{L}})-S(\mathbf{q},+\omega,T_{\mathrm{H}})\\-\frac{1+n(\omega)}{1+n(-\omega)}&[S(\mathbf{q},-\omega,T_{\mathrm{L}})-S(\mathbf{q},-\omega,T_{\mathrm{H}})],\\
S(\mathbf{q},+\omega,T_{\mathrm{H}})=&S(\mathbf{q},-\omega,T_{\mathrm{L}})-S(\mathbf{q},-\omega,T_{\mathrm{H}}).
\end{split}
\end{equation}
As is detailed in the Supplementary Information, the largest parasitic background in our measured inelastic data arises from elastic brag peaks "bleeding" into the inelastic data due to the finite energy width of the beam; this effect is especially prominent at low energy transfers. However, as the elastic Bragg Peaks are nearly temperature independent in strength, this method yields magnetic scattering dominated $S(\mathbf{q},\omega)$ data sets at two temperatures. Finally, the data underwent $D_6$ symmetrization, yielding sets on which all analyses were performed. Figure \ref{fig:dproc} in the Supplementary information shows these processing steps for the $T$=1.7~K dataset.

Additional data at higher energy transfers were taken with an incident neutron energy of $E_i=$12 meV (see Figure 11 in Extended Data). The aforementioned empty can and CYTOP backgrounds were measured and accounted for. However, the $E_i=$12 meV data were only measured at $T=1.7$ K, thus the background subtraction technique of Helton et al. \cite{Helton2007}, as discussed above, was not performed. 

\subsection{Empirical Structure Factor Modeling}
Fitting the scattering data to the equal-time structure factor of Equation \ref{eq:1} involved treating energy-integrated contours as images and performing weighted linear regressions of $\mathbf{q}$-dependent regressors. This process first involved finding the squared magnetic form factor $|F(\mathbf{q})|^2$ and calculating $f_i(\mathbf{q})$ for the crystal in the HK0-plane for each bond. To match the pixel resolution of the scattering data, we transformed these functions into images of the same dimensions. To enhance the accuracy of this model, we added 5$^\circ$ FWHM azimuthal Gaussian ``blurring'' to these images -- centered at $\mathbf{q}$=0 -- to account for the $\sim$5$^\circ$ crystal coalignment error measured in the elastic scattering.

From here, we employed weighted least squares (WLS) linear regressions, with the weight of each pixel inversely proportional to its squared measurement error. These regressions, with heteroscedasticity-consistent standard errors and assuming independence of $\alpha$ with $\rho_i$ but not between $\rho_i$ terms, yielded the parameter estimates shown in Figure \ref{fig:fig4}. Further details on the mechanics of this WLS approach appear in the Supplementary Information.

\subsection{DMRG Calculations}
%\textcolor{red}{TODO: Hongchen: Details about calculation}%
We employ density-matrix renormalization group (DMRG) \cite{White1992} to study the ground state properties of the kagome lattice $J_1$-$J_2$ Heisenberg model with Dzyaloshinskii-Moriya (DM) interaction which is defined by Equation \ref{Eq:Ham} in the main text. Here, $\vec{S}_i$ is the spin-1/2 operator on site $i$ and the first term denotes spin couplings between first-nearest ($J_1$) and second-nearest ($J_2$) neighboring sites, respectively. The DM interaction originates from relativistic spin-orbit coupling and is nonzero when lattice inversion symmetry is absent. In this study, we focus on the $z$-axis component of the DM interaction $\mathbf{D}_{ij}=D_z\hat{z}$ using the convention defined in Lee et al. \cite{Lee2013}. %For simplicity, we have neglected the in-plane component of the DM interaction as it is smaller than the $z$-axis component in the materials of interests \cite{Shekhtman1992,Cepas2008}. %In the DMRG calculation, we set $J_1=1$ as the unit of ground state energy.

In the present study, we take the lattice geometry to be cylindrical with periodic and open boundary conditions in the $\bold{e}_1$ and $\bold{e}_2$ directions, respectively. Here, $\bold{e}_1=(1,0)$ and $\bold{e}_2=(1/2,\sqrt{3}/2)$ (in term of unit cells) denote the two basis vectors of the kagome lattice. We focus on cylinders with width $W$ and length $L$, where $L$ and $W$ are the number of unit cells (and $2L$ and $2W$ are the number of sites) along the $\bold{e}_1$ and $\bold{e}_2$ directions, respectively. Following Yan et al. \cite{Yan2011}, we refer the cylinders to as $\rm YC-2W$, with total number of sites $N=3\times L\times W$. For a natural connection to two dimensions, we consider ``square-like" cylinders $\rm YC$-8 and $\rm YC$-12 with width $W=4 - 6$ and length $L=2W$. We perform up to 50 sweeps and keep up to $m=8000$ states with typical truncation errors $\epsilon\approx 7\times 10^{-7}$ for $\rm YC$-8 and $\epsilon\approx 5\times 10^{-6}$ for $\rm YC$-12 cylinders.

\subsection{Gap Consistency Modeling}
As described in the main text, $\chi''(\omega)$ cuts integrated over varying $\mathbf{q}$-regions suggest possible gap-like behavior in the QSL phase. The argument shown in panel (a) of Figure \ref{fig:fig5} involves modeling $\chi''(\omega)$ curves as a weighted sums of one DHO and one sigmoid component. These components take forms
\begin{equation}\label{eq:GapFit}
\begin{split}
f_{\mathrm{DHO}}(\omega;\omega_0, \Gamma)=\frac{2\omega\Gamma}{(\omega^2_0-\omega^2)^2+\omega^2\Gamma^2},\\ f_{\mathrm{Sigmoid}}(\omega ;a,b)=\left[1+\exp{\left[-b(x-a)\right]}\right]^{-1}.
\end{split}
\end{equation}
Fits were performed using the SciPy package function \texttt{scipy.optimize.curve\_fit}. Common $\omega_0$ and $\Gamma$ values are found for the curves along with independent $a$ and $b$ values and weights sigmoid weights. $a$ values give estimates of gap energies for each curve. For Figure \ref{fig:chippGap} in Extended Data, independent DHO and sigmoid components are used.

\newpage

%%%%%%%%%%%%%%%%%
% Extended Data %
%%%%%%%%%%%%%%%%%

\onecolumngrid
\section*{Extended Data}

\begin{figure*}[h]
\includegraphics[width=0.95\textwidth]{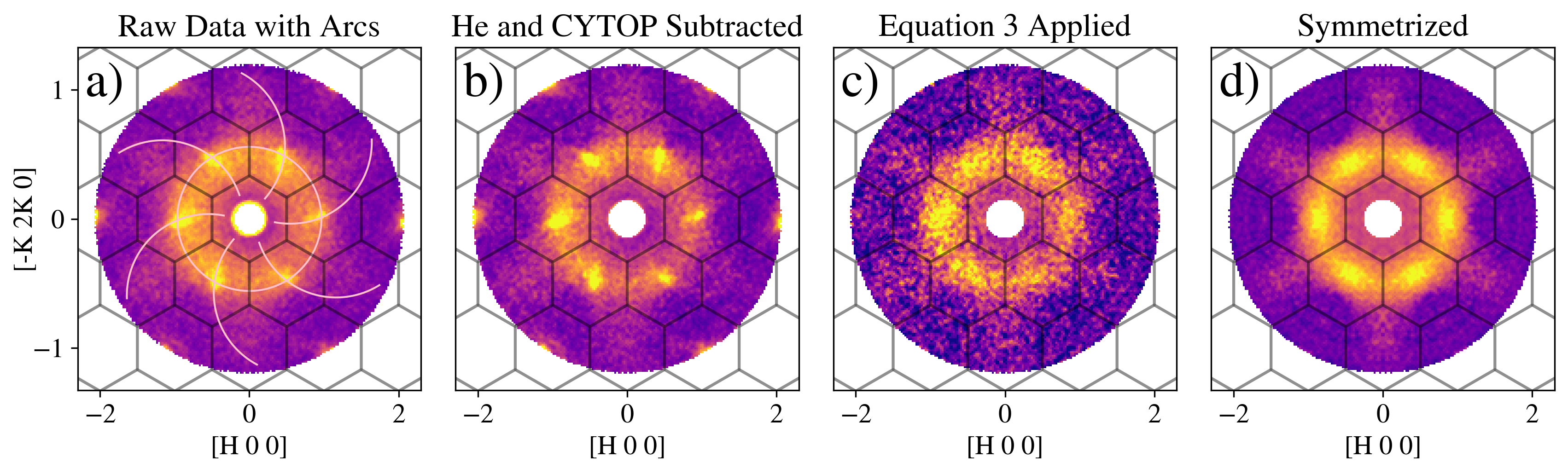}
\caption{\label{fig:dproc} Steps to process inelastic neutron scattering for analysis. (a) Raw scattering data at $\hbar\omega$=[0.3,~0.5]~meV from sample at $T$=1.7~K with incident neutron energy $E_i$=3.32~meV. Parasitic arc-like chiral features stemming from elastic line Bragg peak bleed into the inelastic data are highlighted at the intersection of the light-pink curves overlayed on the data. Details on this elastic bleed effect and its tracking are included in the Supplemental Information. (b) Data after subtraction of He and CYTOP backgrounds with an empirically-determined shielding factor. (c) Removal of temperature-independent background using process of Equation \ref{eq:Helton}. (d) Data after $D_6$ symmetrization, yielding sets on which analyses are performed.}
\end{figure*}

\begin{figure*}[h]
\includegraphics[width=0.95\textwidth]{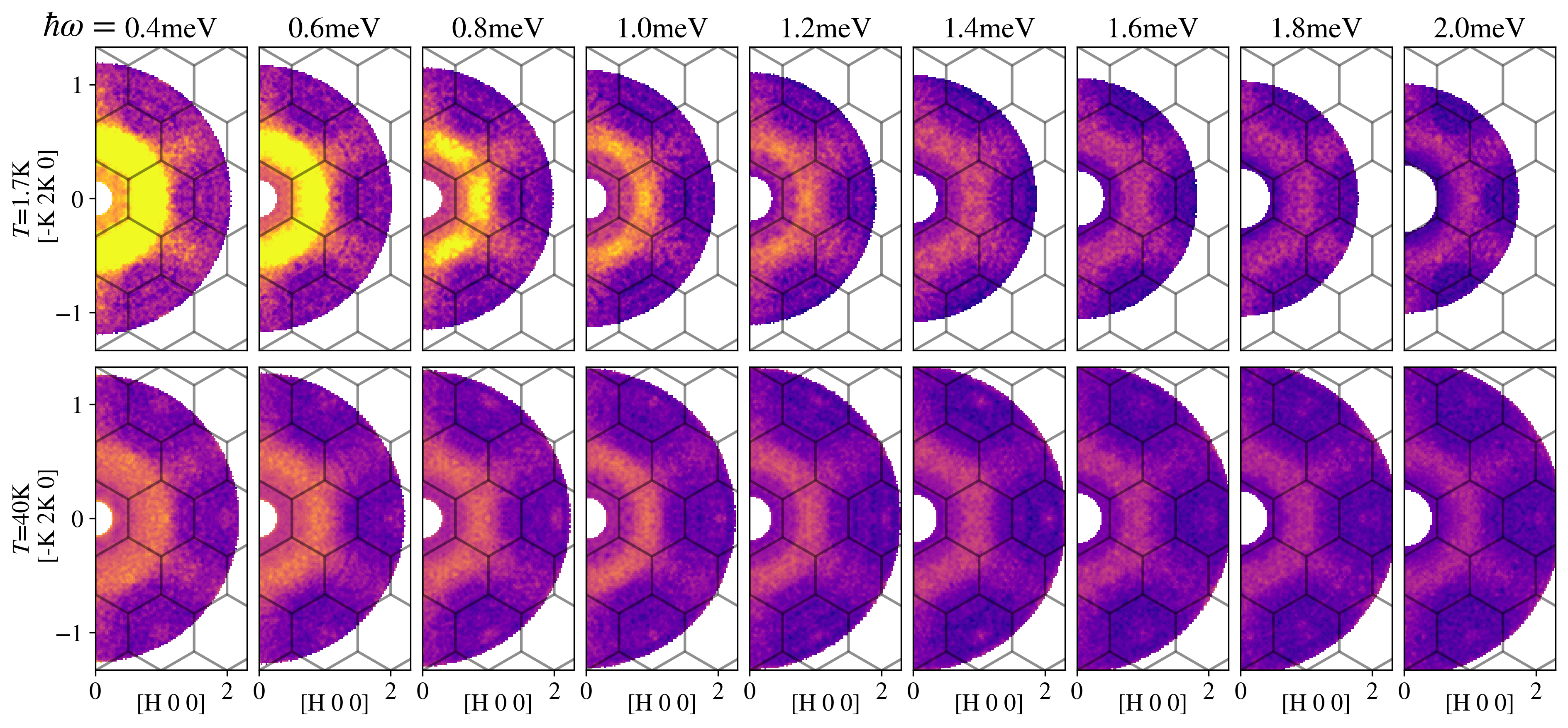}
\caption{\label{fig:figExpo} Scattering contours at various scattering energies for samples at $T$=1.7~K and $T$=40~K. All plots are on the same scale using the same color maps as those in Figures \ref{fig:fig1}-\ref{fig:fig3}.} 
\end{figure*}

\begin{figure*}[h]
\includegraphics[width=0.75\textwidth]{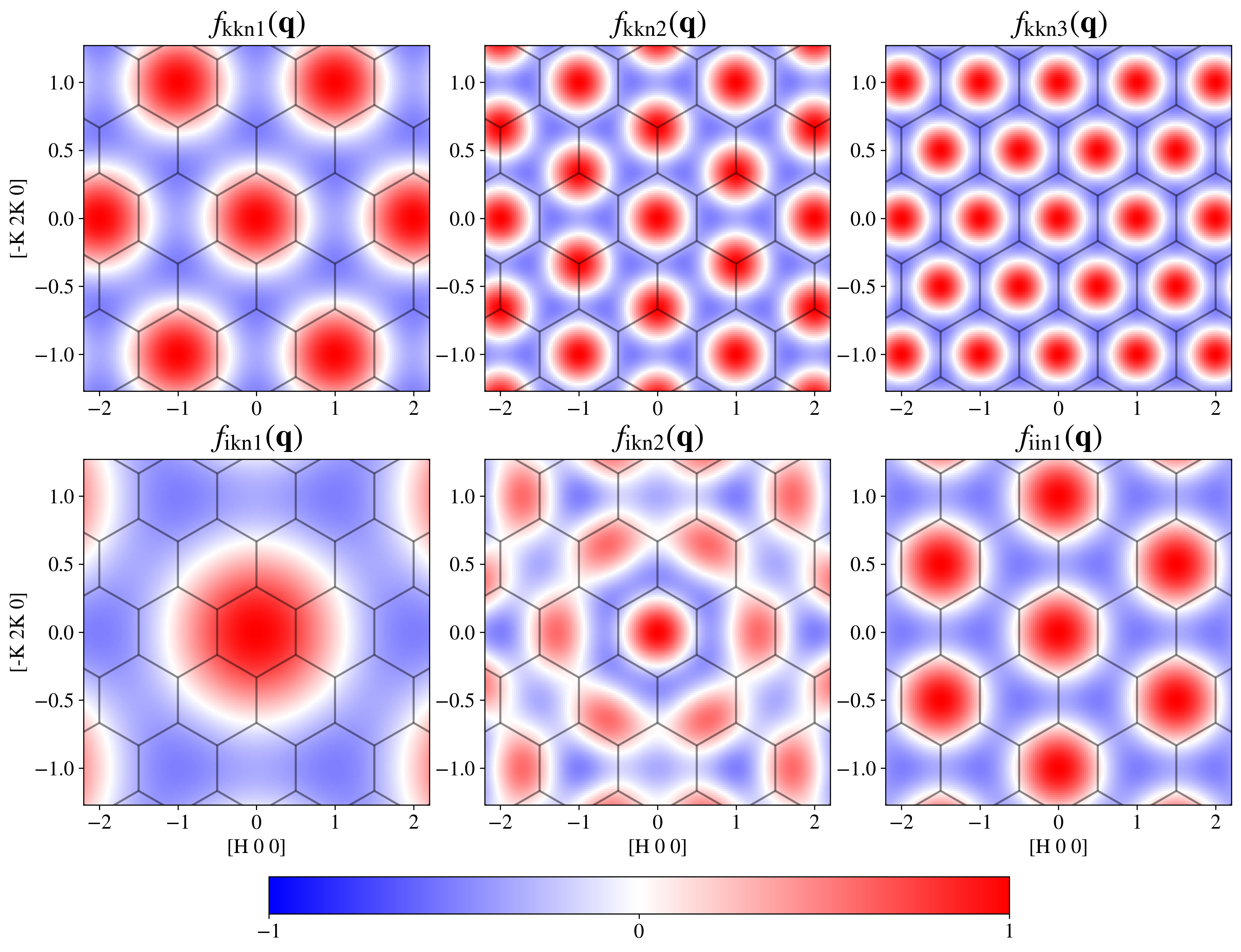}
\caption{\label{fig:qmaps} Components $f(\mathbf{q})$ from Equation \ref{eq:1} for bonds shown in panel (a) of Figure \ref{fig:fig4}. These components weighted and multiplied by the magnetic form factor yield the model shown in panels (a, c) in Figure \ref{fig:fig1}.}
\end{figure*}

\begin{table*}[h]
{\renewcommand{\arraystretch}{2}%
\begin{tabular}{|l|l|l|l|l|l|l|l|}
\hline
$\hbar\omega$ (meV) & Material        & kkn1 & kkn2 & kkn3 & ikn1 & ikn2  & iin1  \\ \hline
\multirow{2}{*}{    0.4} & Zn-barlowite    & -0.26 & 0.04 & 0.13 & 0.35 & -0.25 & -0.18 \\ \cline{2-8} & herbertsmithite & -0.12 & 0.00 & 0.01 & 0.10 & -0.04 & -0.24 \\ \hline
\multirow{2}{*}{        1.3} & Zn-barlowite    & -0.38 & 0.07 & 0.08 & 0.12 & -0.14 & -0.12 \\ \cline{2-8} & herbertsmithite & -0.22 & 0.16 & 0.14 & 0.06 & 0.04  & -0.13 \\ \hline
\end{tabular}}
\caption{\label{tab:compths} Fitted $\rho_i$ values using model from Figures \ref{fig:fig1} and \ref{fig:fig4} on scattering data from Zn-barlowite and herbertsmithite with scattering energies $\hbar\omega$=0.4~meV and $\hbar\omega$=1.3~meV. Bonds are labeled according to the convention introduced in Figure \ref{fig:fig4}(a). Both the signs of the correlation signals and their relative changes between the two energies are remarkably consistent between the materials.}
\end{table*}

\begin{figure*}[h]
\includegraphics[width=0.55\textwidth]{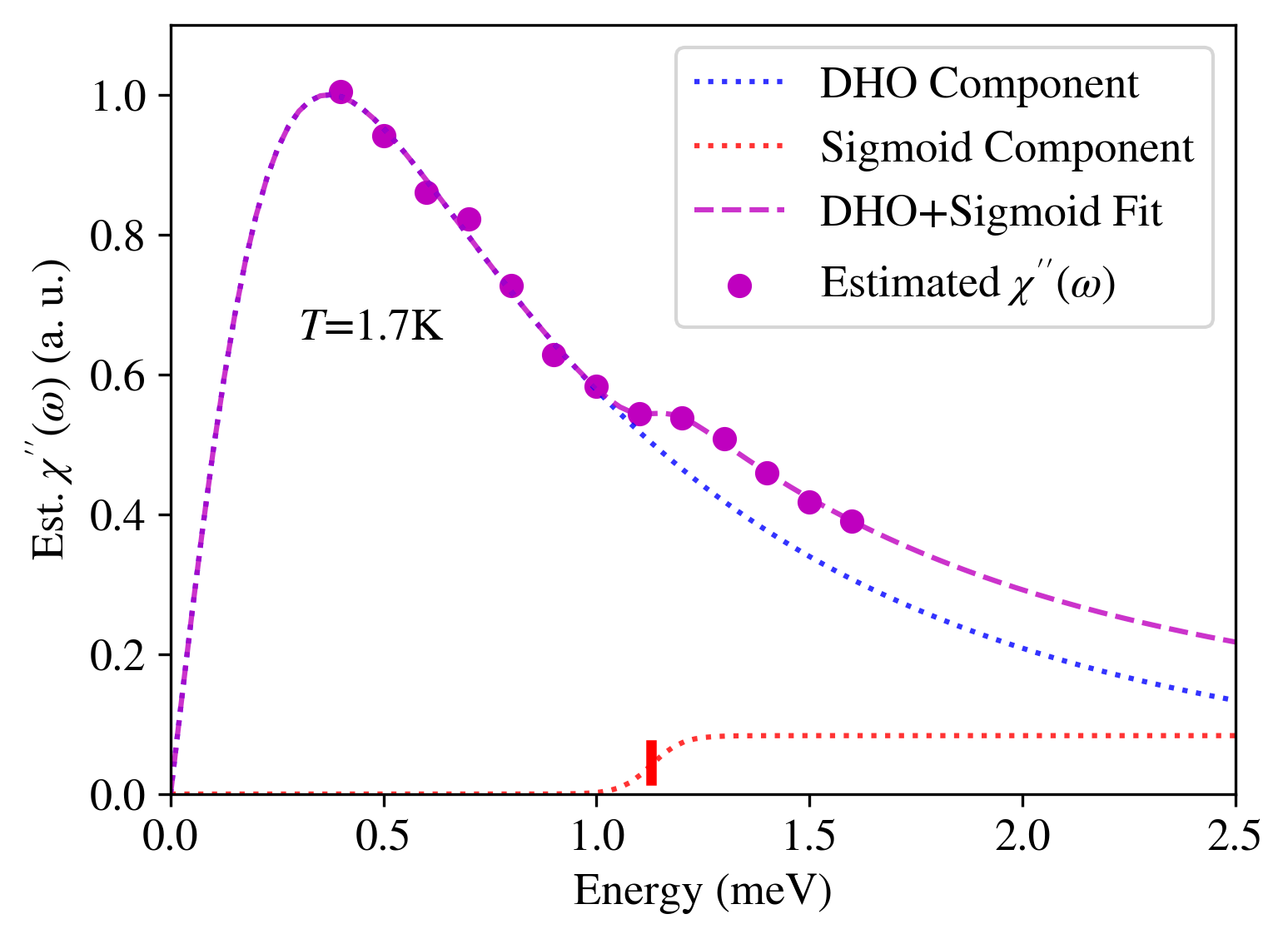}
\caption{\label{fig:chippGap} Fits to estimated $\chi''(\omega)$ from $T$=1.7~K using Equation \ref{eq:GapFit}. Using independent DHO and sigmoid components, there appears to be a signature consistent with a $\sim$1.1~meV energy gap. Note that the estimated $\chi''(\omega)$ derives from removing $q$-dependent signals from bond correlations and is the best estimate of $\int d\mathbf{q}\:\chi''(\mathbf{q},\omega)$. Hence, this indicates the existence of an ``average'' gap over all $\mathbf{q}$.}
\end{figure*}

\begin{figure*}[h]
\includegraphics[width=0.85\textwidth]{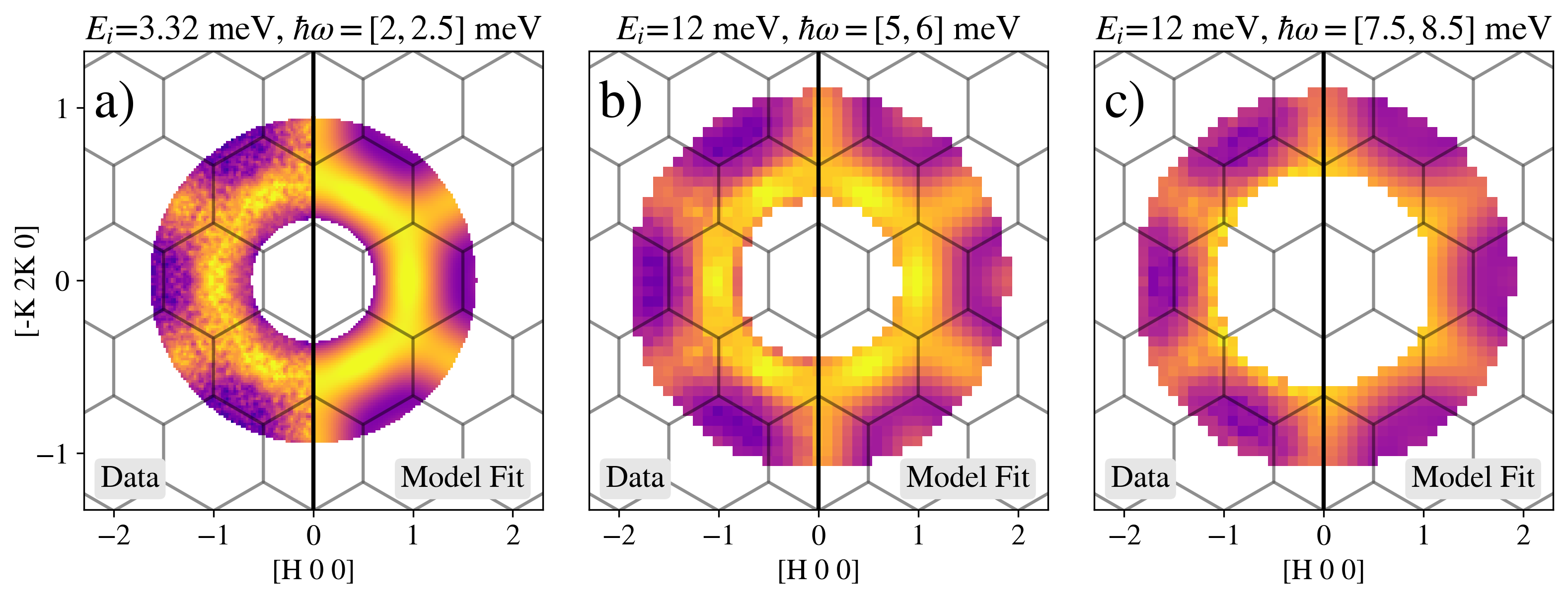}
\caption{\label{fig:highEnergy} Higher energy inelastic scattering at $T$=1.7~K (using measurements at a higher incident neutron energy $E_i$=12~meV) with correlation model comparisons. (a) Data measured with $E_i$=3.32~meV and integrated over $\hbar\omega = [2.0, 2.5]$~meV. (b, c) Contours measured with $E_i$=12~meV, integrated over $\hbar\omega = [4.5, 5.5]$~meV and $\hbar\omega = [7.5, 8.5]$~meV, respectively; these lack detail temperature-independent background subtraction. Fits with the model from Fig. 4 for data in (a) yield $\rho_i$ values [kkn1, kkn2, kkn3] of [-0.51(2), 0.19(1), 0.25(1)]. For data in (b, c) a model using only three nearest kagome layer correlations [kkn1, kkn2, kkn3] respectively yield  [-0.24(2), 0.13(1), 0.19(2)] and [-0.30(1), 0.11(1), 0.08(2)]. The measured dynamic structure factor from 2 meV up to 8.5 meV is consistent with kagome-kagome pairs with AF nearest neighbor correlations and weaker FM 2nd and 3rd nearest nearest neighbor correlations, consistent with the fits in Fig 4.}
\end{figure*}

\end{document}

\preprint{APS/123-QED}

%\title{Manuscript Title:\\with Forced Linebreak}% Force line breaks with \\
\title{Supplementary Material for: Identifying Universal Spin Excitations in Spin-1/2 Kagome Quantum Spin Liquid Materials}
%\thanks{A footnote to the article title}%

\author{Aaron T. Breidenbach}
\altaffiliation[]{These authors contributed equally to this work}
\affiliation{Stanford Institute for Materials and Energy Sciences, SLAC National Accelerator Laboratory, Menlo Park, CA 94025, USA}
\affiliation{Department of Physics, Stanford University, Stanford, CA 94305, USA}
\author{Arthur C. Campello}
\altaffiliation[]{These authors contributed equally to this work}
\affiliation{Stanford Institute for Materials and Energy Sciences, SLAC National Accelerator Laboratory, Menlo Park, CA 94025, USA}
\affiliation{Department of Applied Physics, Stanford University, Stanford, CA 94305, USA}
\author{Jiajia Wen}
\author{Hong-Chen Jiang}
\affiliation{Stanford Institute for Materials and Energy Sciences, SLAC National Accelerator Laboratory, Menlo Park, CA 94025, USA}
\author{Daniel M. Pajerowski}
\affiliation{Neutron Scattering Division, Oak Ridge National Laboratory, Oak Ridge TN 37830, USA}
\author{Rebecca W. Smaha}
\affiliation{Stanford Institute for Materials and Energy Sciences, SLAC National Accelerator Laboratory, Menlo Park, CA 94025, USA}
\affiliation{Department of Chemistry, Stanford University, Stanford, CA 94305, USA}
\altaffiliation[Present address: ]{National Renewable Energy Laboratory, Golden, CO 80401, USA}
\author{Young S. Lee}
\email{youngsl@stanford.edu}
\affiliation{Stanford Institute for Materials and Energy Sciences, SLAC National Accelerator Laboratory, Menlo Park, CA 94025, USA}
\affiliation{Department of Applied Physics, Stanford University, Stanford, CA 94305, USA} 
\date{\today}
\maketitle

\onecolumngrid

\section*{Shielding Factor Determination}

The Methods section mentions that to accurately subtract background contributions from Helium and CYTOP from our data sets, we empirically estimate a shielding factor parameter. Here, we define this parameter in more detail and explain precisely how we estimate it to derive our processed data. For each temperature at which we measured data, we made three measurements: One on the entire sample assembly -- with the crystals, CYTOP adhesive, and Helium in the background, one on a ``dummy'' sample -- a sample without crystals but with a nearly matching amount of CYTOP -- and background Helium, and one with no sample in the beam path and only Helium. For clarity, we call these measurements, respectively, $\mathcal{M}[\mathrm{Sample + He + CYTOP}]$, $\mathcal{M}[\mathrm{He + CYTOP}]$, and $\mathcal{M}[\mathrm{He}]$. We also let $\mathcal{D}[\mathrm{Sample}]$ be our best estimate of the data only from the crystal sample.

In theory, one would imagine that subtracting $\mathcal{M}[\mathrm{He + CYTOP}]$ from $\mathcal{M}[\mathrm{Sample + He + CYTOP}]$ would yield a good estimate of $\mathcal{D}[\mathrm{Sample}]$. Unfortunately, this approach does not account for the effect of neutrons being shielded from the CYTOP by the sample itself. For this reason, practical background subtraction for all temperatures is best approached with the model  
\begin{equation}
\mathcal{D}[\mathrm{Sample}] = \mathcal{M}[\mathrm{Sample + He + CYTOP}] - \beta\: \mathcal{M}[\mathrm{He + CYTOP}] - (1-\beta)\mathcal{M}[\mathrm{He}],
\label{eq:shieldF}
\end{equation}
where $\beta$ is a ``self-shielding factor'' parameter. Note that for all choices of $\beta$, the He signal is fully accounted for. This only changes the amount of the CYTOP-only signal that is subtracted from $\mathcal{M}[\mathrm{Sample + He + CYTOP}]$ to estimate $\mathcal{D}[\mathrm{Sample}]$.

\begin{figure*}[h]
\includegraphics[width=0.5\textwidth]{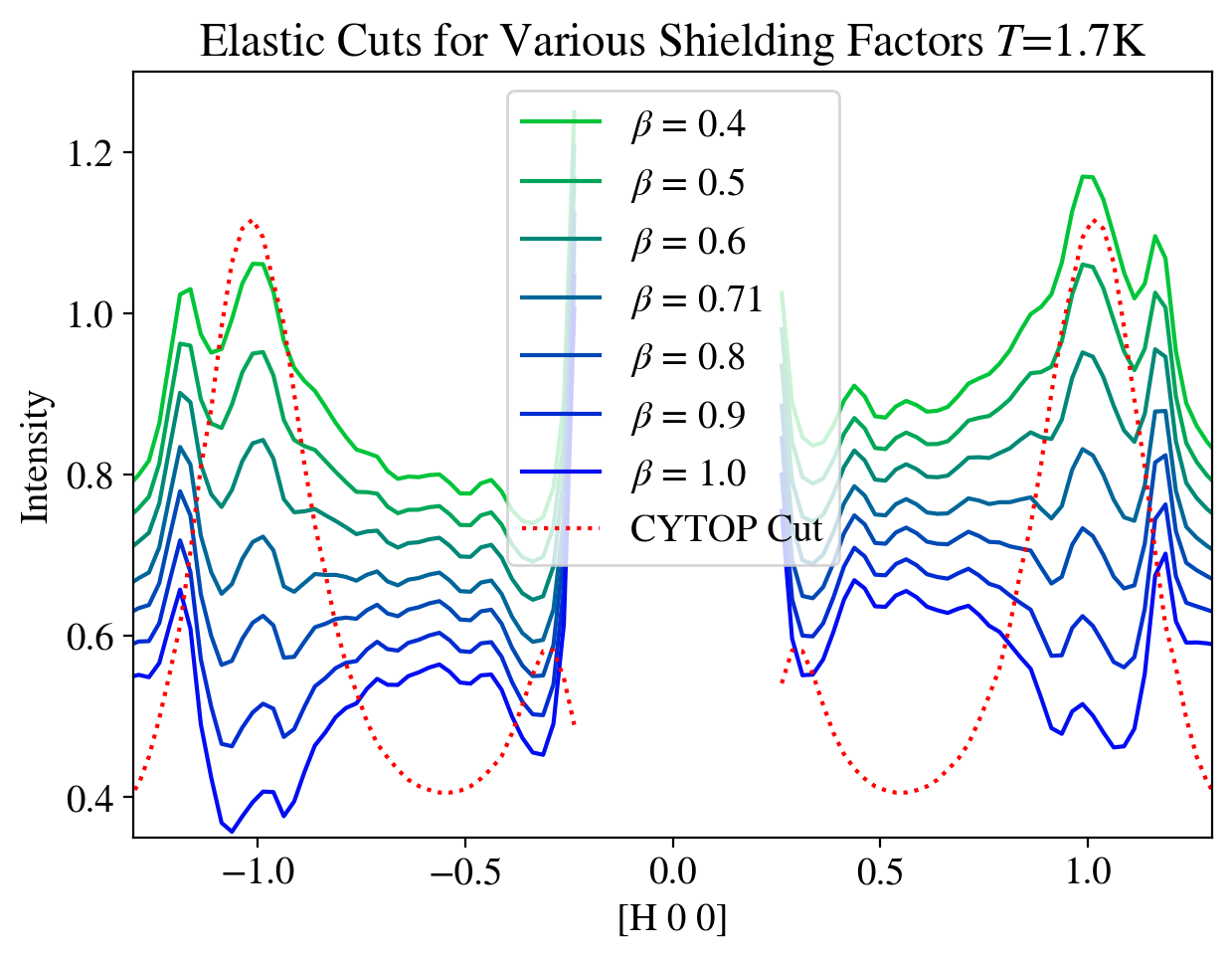}
\caption{\label{fig:sfact} Averaged cuts along [H 0 0] and equivalent directions of the elastic scattering pattern -- $\hbar\omega$ = [-0.1, 0.1]~meV -- for data processed with various shielding factors using Equation \ref{eq:shieldF} applied to $T$=1.7~K data. Red dotted line shows this same cut for measured [CYTOP + He] data. The signature of the CYTOP is clear in the elastic data and a shielding factor choice of 0.71 appears optimal.  }
\end{figure*}

Empirically, we find that the CYTOP signal is strongest in the elastic line: i.e. where there is zero energy transfer. Figure \ref{fig:sfact} shows this elastic signal from CYTOP integrated along the [H 0 0] and equivalent directions measured at $T$=1.7~K. It also shows $\mathcal{D}[\mathrm{Sample}]$ estimated with various shielding factors.
It is clear that in the $\beta=0$ case the CYTOP is under-subtracted -- as it is ignored completely -- but in the $\beta=1$ case there is over-subtraction of the CYTOP signal. By examining plots like these, we found choices for $\beta$ that minimized CYTOP signals for all data sets. For $T$=1.7~K and $T$=40~K data, the best choice was clearly $\beta$=0.71. The noisiness of the $T$=150~K data made this parameter much harder to estimate, but it was chosen to be $\beta=1.75$, although all values in the range $\beta \in [1.5, 2.0]$ seemed reasonable. Note that the dummy sample for the 150K background was taken in a slightly different geometry, which is why we have $\beta>1$; normally physical constraints would govern $0<\beta<1$ for a dummy sample in the same geometry and the same amount of CYTOP. The figure below reproduces Figure 3 with the $T$=150~K cut reflecting this shielding factor uncertainty, which results in higher error bars.
\begin{figure*}[h]
\includegraphics[width=0.5\textwidth]{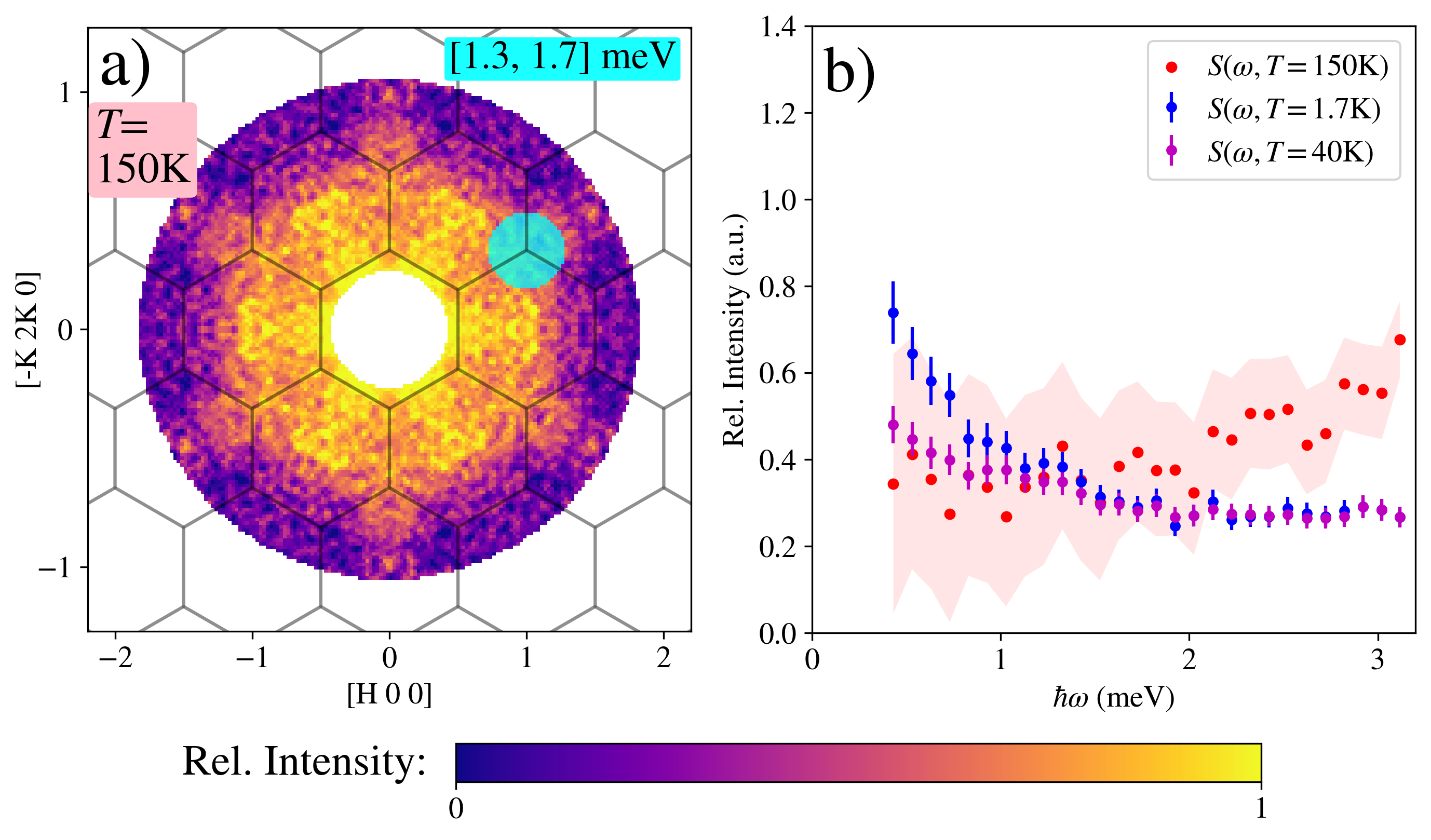}
\caption{\label{fig:150kerr} Figure 3 from the main text recreated with $T$=150~K data cut showing errors reflecting shielding factor uncertainties.}
\end{figure*}

\section*{Origin of Chiral Detector Artifacts in Unprocessed Scattering Data}

Figure 9 in Extended Data shows the steps of processing the incoming CNCS data. The raw data in panel (a) shows strong scattering features at [1 0 0] and equivalent positions. These appear as ``streaks'' of a chiral nature and persist with subtractions of Helium and CYTOP backgrounds (regardless of shielding factor) as shown in panel (b). Panel (c) shows that this artifact disappears with the detail-balance background subtraction outlined in Equation 5 in the Methods section of the main text. Here we show the origins of this artifact, model how it evolves with scattering energy, and explain how our background subtraction removes it. Because the sample is made from a collection of crystals, any intrinsic chiral scattering feature would appear with $D_6$ symmetry. Hence, the streaks cannot be intrinsic, since there is no corresponding steak with the opposite chirality. As we derive below, they ultimately arise from the finite energy width of the incident neutron beam. Though the width of the incident energy distribution is $\sim$0.1~meV, the weakness of the diffuse magnetic scattering relative to that of elastic scattering means the artifact persists well into the inelastic signal for our data.

To track this effect, we consider the time-of-flight scattering geometry of the CNCS instrument and the effects of the energy width of the incident beam. For a normal inelastic scattering event from an incoming beam of energy $E_i$, the detected $\mathbf{q}$ in units $[\mathrm{\AA}^{-1}]$ is 
\begin{equation}
\mathbf{q}=\mathbf{k}_f-\mathbf{k}_i=k(E_i)\left[ \left( \sqrt{1-\frac{\Delta E}{E_i}}\cos\phi -1 \right) \hat{i} +\left( \sqrt{1-\frac{\Delta E}{E_i}}\sin\phi -1 \right) \hat{j} \right],
\label{eq:arc1}
\end{equation}
where $\Delta E$ is the detected loss in energy from the time of flight, $k(E)$ is the incident beam wavenumber, and $\phi$ is the angle between initial and final wave vectors (and equal to twice the Bragg angle, $\phi=2\theta_B$). We now suppose the incident energy $E_i$ is actually $E_i+\delta$, where $\delta$ is some deviation that may be positive or negative. From the time of flight geometry, we can calculate the effect of this deviation $\delta$ on the detected $\Delta E$ from an elastic scattering event according to the function
\begin{equation}
\Delta E_d(\delta) = E_i\left[\left(\frac{d_1}{d_2}-\frac{d_1+d_2}{d_2}\sqrt{\frac{E_i}{E_i+\delta}}\right)^{-2}-1\right].
\label{eq:arc2}
\end{equation}
At the CNCS instrument, $d_1$=36.25~m and $d_2$=3.75~m. Figure \ref{fig:arc} below shows this relationship and the FWHM of $\delta$ given our chosen incident energy and chopper speed. Note that it is nearly linear in our regime of interest.
\begin{figure*}[h]
\includegraphics[width=0.5\textwidth]{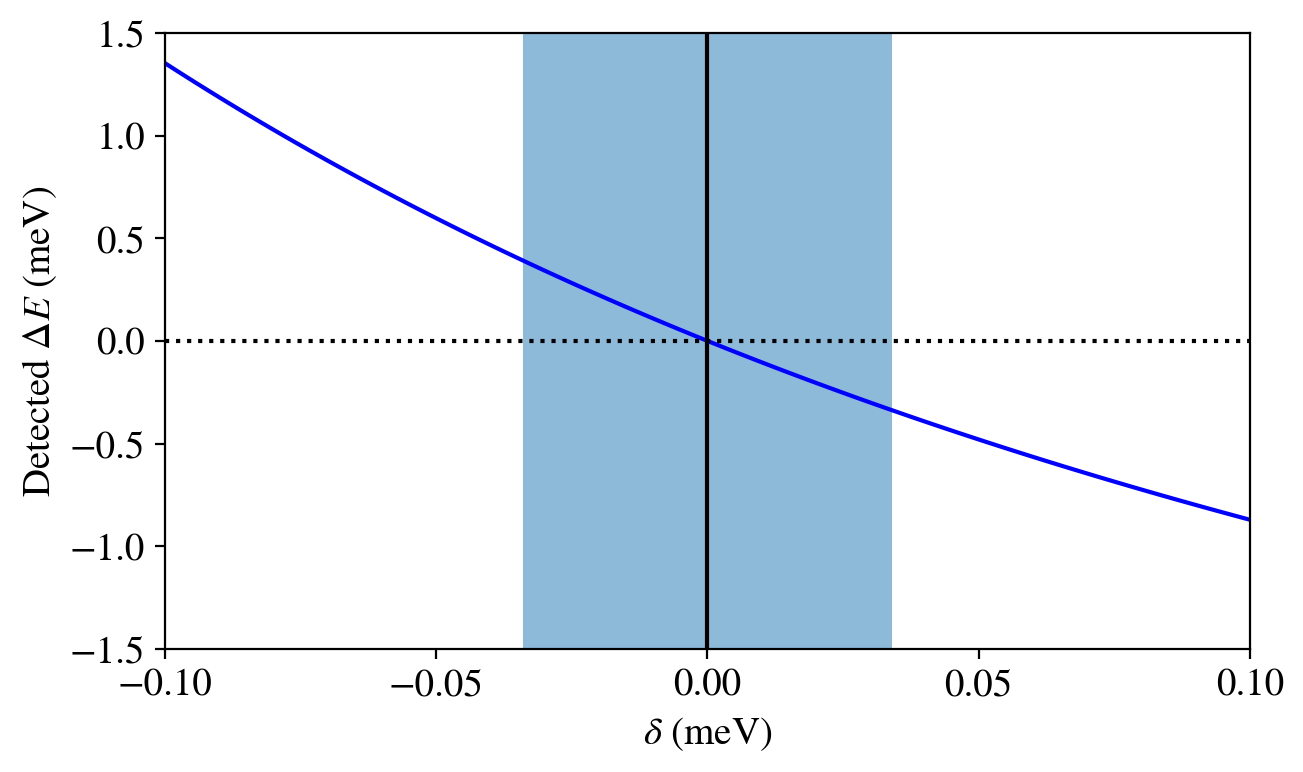}
\caption{\label{fig:arc} Detected $\Delta E$ as a function of incident neutron energy deviation $\delta$ according to Equation \ref{eq:arc2} given CNCS instrument geometry and incident neutron energy $E_i$=3.32~meV. The shaded region shows the reported FWHM for $\delta$ at the CNCS for the 180Hz DD-chopper frequency used in our experiments.}
\end{figure*}

Alone, Equation \ref{eq:arc2} accounts for the ``bleeding'' of elastic Bragg features into the inelastic signal, but does not explain the chiral nature of the artifacts. This quality comes from the combination of Equations 2 and 3, which shows that a non-zero $\delta$ also affects the detected $\mathbf{q}$ position of elastic scattering events. Specifically, an elastic Bragg scattering event that should be detected at 
\begin{equation}
\mathbf{q}=k(E_i)\left[\left(\cos\phi-1\right)\hat{i}+(\sin\phi-1) \hat{j} \right]
\label{eq:arc3}
\end{equation}
is instead detected at
\begin{equation}
\mathbf{q}_d=k(E_i+\delta)\left[ \left( \sqrt{1-\frac{\Delta E_d(\delta)}{E_i+\delta}}\cos\phi -1 \right) \hat{i} +\left( \sqrt{1-\frac{\Delta E_d(\delta)}{E_i+\delta}}\sin\phi -1 \right) \hat{j} \right].
\label{eq:arc4}
\end{equation}
Using Equations \ref{eq:arc2} and \ref{eq:arc4}, one can model and track how this process contaminates the detected inelastic data. This is done in Panel (a) of Figure 9 for extraneous scattering from [1 0 0] and equivalent Bragg peaks at $\hbar\omega = [0.3, 0.5]$~meV. The intersection of the $\mathbf{q}=0$ centered circle and the arcs show where the strongest signal from Bragg peaks themselves appears and the arcs follow the streaking caused by the variance in $\delta$. In the figure chiral artifacts from [2 0 0] equivalent Bragg peaks also appear, but these are not tracked. 

Thankfully, this explained artifact can be removed with the collection of two datasets at different temperatures and the use of the background subtraction method detailed in the Methods section. Relative to most scattering processes measured by neutrons, this effect is small in intensity. It only appears in this work because of the weak and diffuse nature of the signal we measure in our sample.

\section*{Structure Factor Modeling and Weighted Least Squares Estimation}

The model described in the main text estimates the momentum and energy dependent scattering according to the form 
\begin{eqnarray} 
S_{\mathrm{mag}}(\mathbf{q}, \omega)\approx \alpha(\omega) |F(\mathbf{q})|^2\left(1+2\sum^n_{i=1}\rho_i(\omega) f_i(\mathbf{q})\right),
\label{eq:1}
\end{eqnarray}
parameterized by $\alpha(\omega)$ and $\rho_i(\omega)$ for each $i$ bond. This model derives from the static structure factor equation adapted from Han et al. \cite{Han2016}
\begin{eqnarray}
S_{\mathrm{mag}}(\mathbf{q})=  N\langle \vec{S}^2\rangle |F(\mathbf{q})|^2\left(1+2\sum^n_{i=1}\frac{m_i\langle \vec{S}\vec{S}'\rangle_i}{N\langle \vec{S}^2\rangle}f_i(\mathbf{q})\right).
\end{eqnarray}
Here, $\alpha(\omega)$ gives the energy dependence of the $N\langle \vec{S}^2\rangle$, the amount of scattering whose $\mathbf{q}$-dependence is $|F(\mathbf{q})|^2$. For each bond included, the regression yields a term $\rho_i(\omega)= (m_1/N)\left(\langle \vec{S}\vec{S}'\rangle_i/\langle \vec{S}^2\rangle\right)$. The coefficient $(N\langle \vec{S}^2\rangle)^{-1}$ in its definition means the magnitude of $\rho_i(\omega)$ represents the relative amount of scattering that with a $\mathbf{q}$-dependence of $f_i(\mathbf{q})$. Hence, it is best described as a ``relative correlation signal.'' Though is $\rho_i(\omega)$ proportional to the energy-dependent spin correlation $\langle \vec{S}\vec{S}'\rangle_i/\langle \vec{S}^2\rangle$, comparing spin correlations across bonds using $\rho_i(\omega)$ requires weighting by $m_i/N$, the energy-independent relative bond density. These bond densities are easy to calculate for bonds intrinsic to the kagome lattice as there is no disorder over those bonds. In the case of impurities, calculating these weights requires assumptions of impurity abundance and distribution. As stated in the main text, we use the data to derive estimates for these parameters with appropriate errors. Interpreting these parameters requires understanding their meanings and limitations.

We now include a mathematically complete explanation of the formalism behind the derived parameter estimates and standard errors found in Figure 4 (b-e) of the main text. Starting from the very basics, we note that the detector at the CNCS instrument obtains scattered intensity data ``events'' wherein a detected neutron's energy and two-dimensional momenta are recorded. Later processing bins this data into a three-dimensional histogram over discrete $(q_1,q_2,\omega)$ coordinates. Integrating these histograms over energy windows $[\omega-\delta\omega,\omega+\delta\omega]$ allows us to write Equation 3 of the main text using matrices as
% image-form empirical structure factor model
\begin{equation}\label{eq:ifesfm}
\mathcal{I}_{[\omega-\delta\omega,\omega+\delta\omega]}\approx\alpha^{(\omega)}A_{F}\circ\left(J+ 2\sum^n_{i=1}\rho^{(\omega)}_iA_i\right)
\end{equation}
where $A_F$ is the squared magnetic form factor image, $A_i$ is the image corresponding to $f_i(\mathbf{q})$, and $J$ is the all-ones matrix -- note that $\circ$ is the element-wise multiplication operation. Since Equation \ref{eq:ifesfm} is a linear regression and the errors of measured ``pixels'' are known to follow uncorrelated counting statistics, we approach the minimization using weighted least squares (WLS) formalism. We first let $\mathcal{M}_{[\omega-\delta\omega,\omega+\delta\omega]}$ represent the measured image. Then, we unravel the image matrix $A$ and index the $j$th of $k$ pixels as $(A)_j$. Letting $A'_i=A_F\circ A_i$ we define
% defining regression matrices and vectors
\begin{equation}\label{eq:drmv}
\mathbf{y}^{(\omega)} = 
\begin{bmatrix}
\left(\mathcal{M}_{[\omega-\delta\omega,\omega+\delta\omega]}\right)_1 \\
\vdots \\
\left(\mathcal{M}_{[\omega-
\delta\omega,\omega+\delta\omega]}\right)_k
\end{bmatrix},\quad 
X =
\begin{bmatrix}
(A_F)_1 & (A'_1)_1&\dots & (A'_n)_1 \\
(A_F)_2 & (A'_1)_2&\dots & (A'_n)_2 \\
\vdots & \vdots & \ddots & \vdots \\
(A_F)_k & (A'_1)_k&\dots & (A'_n)_k
\end{bmatrix},\quad 
\mathbf{\beta}^{(\omega)} = 
\begin{bmatrix}
\alpha^{(\omega)} \\
2\alpha^{(\omega)}\rho^{(\omega)}_1 \\
\vdots \\
2\alpha^{(\omega)}\rho^{(\omega)}_n
\end{bmatrix},
\end{equation}
and weights $\mathbf{w}^{(\omega)}$ such that $w^{(\omega)}_j=1\big/\sigma^2\!\left(\mathcal{M}_{[\omega-\delta\omega,\omega+\delta\omega]}\right)_j$, i.e. the inverse variance of the integrated slice at the $j$th pixel. Letting $W^{(\omega)}=\mathrm{diag}\left(\mathbf{w}^{(\omega)}\right)$ yields the WLS estimate
% WLS structure factor slice
\begin{equation}\label{eq:wlssfs}
\hat{\beta}^{(\omega)}_{\mathrm{WLS}}=\left(X^\intercal W^{(\omega)}X\right)^{-1}X^\intercal W^{(\omega)}\mathbf{y}^{(\omega)}\implies \hat{\rho}^{(\omega)}_i = \frac{1}{2}
\frac{\left(\hat{\beta}^{(\omega)}_{\mathrm{WLS}}\right)_{2+i}}{\left(\hat{\beta}^{(\omega)}_{\mathrm{WLS}}\right)_{1}}.
\end{equation}
The known errors of the residuals allow one to find the standard errors of the estimate $\hat{\beta}^{(\omega)}_{\mathrm{WLS}}$ to be 
% WLS structure factor slice error
\begin{equation}\label{eq:wlssfse}
\mathrm{Var}\left(\hat{\beta}^{(\omega)}_{\mathrm{WLS}}\right)= \left(X^\intercal X\right)^{-1}\left(X^\intercal \left[W^{(\omega)}\right]^{-1}X\right)\left(X^\intercal X\right)^{-1},\qquad SE\left(\hat{\beta}^{(\omega)}_{\mathrm{WLS}}\right)_i = \sqrt{\mathrm{Var}\left(\hat{\beta}^{(\omega)}_{\mathrm{WLS}}\right)_{ii}}.
\end{equation}
Letting $\left({\mathbf{\sigma}^2}^{(\omega)}_{\mathrm{WLS}}\right)_i=\mathrm{Var}\left(\hat{\beta}^{(\omega)}_{\mathrm{WLS}}\right)_{ii}$ and cautiously assuming independence between $\alpha$ and $\rho_i$, we find standard errors for $\rho_i$ via the propagation
% WLS correlation error
\begin{equation}\label{eq:wlsce}
SE\left(\hat{\rho}^{(\omega)}_i\right)\approx\sqrt{\left[\left({\mathbf{\sigma}^2}^{(\omega)}_{\mathrm{WLS}}\right)_1+\left(\hat{\beta}^{(\omega)}_{\mathrm{WLS}}\right)^2_{1}\right]^{-1}\left[\frac{1}{4}\left({\mathbf{\sigma}^2}^{(\omega)}_{\mathrm{WLS}}\right)_{1+i}-\left({\mathbf{\sigma}^2}^{(\omega)}_{\mathrm{WLS}}\right)_1\left(\hat{\beta}^{(\omega)}_{\mathrm{WLS}}\right)^2_{1+i}\right]}.
\end{equation}
Note the use of hats to denote an estimation; this is common notation in fields where linear regressions are more commonly used, like in Econometrics. In the main text, this notation is abandoned and estimates are described as such. This regression approach has been shown to be a heteroskedasticity-consistent estimator \cite{White1980}. This means it produces standard errors that are robust to regressor-dependent variance, which is the case in our models due to the $\mathbf{q}$-dependence of statistical counting errors.

We note that in cases where $\mathbf{q}$-coverage is limited, a phenomenon known as multicollinearity can occur, where high correlations between $f_i(\mathbf{q})$ over the available data lead to large and correlated errors in parameter estimates. This is a well-studied issue in linear regression and amounts to ambiguity of a trade-off between regressors that appear similar. Multicollinearity exacerbates the effects of counting errors on coefficient uncertainties and also makes the matrix inversion in the equation for $\hat{\beta}_{WLS}$ above less numerically stable. For the data presented, at high $\hbar\omega$ the $\mathbf{q}$-coverage diminishes and the counting statistics become worse. Hence, in the main text, we terminate the energy-dependence of our model at energies where the fits are still highly reliable. 

\bibliography{sample}